\documentclass[pdftex,singlecolumn,epjc3]{svjour3} 
\RequirePackage{graphicx}
\RequirePackage{mathptmx}      
\RequirePackage{flushend}
\RequirePackage[numbers,sort&compress]{natbib}
\RequirePackage[colorlinks,citecolor=blue,urlcolor=blue,linkcolor=blue]{hyperref}
\usepackage{amsmath}
\usepackage{graphicx}
\usepackage{geometry}
\usepackage{caption}
\usepackage{subcaption}
\usepackage{color}
\begin{document}
\baselineskip=18 pt
\begin{center}
{\large{\bf Relativistic Motions of Spin-Zero Quantum Oscillator Field in a Global Monopole Space-Time with External Potential and AB-effect }}
\end{center}

\vspace{0.5cm}

\begin{center}
{\bf Faizuddin Ahmed}\\
(\tt faizuddinahmed15@gmail.com ; faizuddin@ustm.ac.in)\\
{\bf Department of Physics, University of Science \& Technology Meghalaya, Ri-Bhoi, Meghalaya-793101, India }
\end{center}

\vspace{0.6cm}

\begin{abstract}:
In this paper, we analyze a spin-zero relativistic quantum oscillator in the presence of the Aharonov-Bohm magnetic flux in a space-time background produced by a point-like global monopole (PGM). Afterwards, we introduce a static Coulomb-type scalar potential and subsequently with the same type of vector potential in the quantum system. We solve the generalized Klein–Gordon oscillator analytically for different functions ({\it e. g.} Coulomb- \& Cornell-type functions) and obtain the bound-states solutions in each case. We discuss the effects of topological defects associated with the scalar curvature of the space-time and the Coulomb-type external potentials on the energy profiles and the wave function of these oscillator fields. Furthermore, we show that the obtained energy eigenvalues depend on the magnetic quantum flux which gives rise to the gravitational analogue of the Aharonov-Bohm (AB) effect. 
\end{abstract}

\vspace{0.3cm}


\vspace{0.3cm}

{\bf PACS Number(s):} 03.65.Pm, 03.65.Ge, 14.80.Hv

\section{Introduction}
\label{intro}

The relativistic quantum motions of scalar and spin-half particles under gravitational effects produced by different curved space-time geometries, for instance, G\"{o}del and G\"{o}del-type space-times \cite{AA1,AA2}, topologically trivial \cite{AA3}, and nontrivial \cite{LCNS} space-times, topological defects (which will discuss in the next paragraphs) have been of interest among researchers. The spin-$0$ scalar particles (bosons) are described by the Klein-Gordon equation while the Dirac equation for the spin-half particles (fermion). These wave equations have been solved using interaction potentials (scalar and vector) of different kinds in quantum systems by different techniques ({\tt e. g.}, power series method, supersymmetric approach, factorization method, the Nikiforov-Uvarov method). Moreover, a harmonic oscillator acts as a prototype model in different areas of physics, such as condensed matter physics, quantum statistical mechanics, and quantum field theory. This quantum field theory in curved space-time is considered the first approximation to quantum gravity. In the gravitational background, it is necessary to analyze a single-particle state to make a consistent quantum field theory. A well-known version for the relativistic harmonic oscillator or additional form of interaction was proposed in Ref. \cite{MM} for a spin-half particle which is called the Dirac oscillator. This new alternative form of interaction furnishes the Schr\"{o}dinger equation with a harmonic oscillator potential in the non-relativistic scheme \cite{SP}. Inspired by this Dirac oscillator model, a similar model for spin-zero particles (bosons) is studied by the substitution of the radial momentum operator $\vec{p} \to (\vec{p}-i\,M\,\omega\,\vec{r})$ in the Klein-Gordon equation, and this new model is called the Klein-Gordon oscillator. Several researchers have been investigated this Klein-Gordon oscillator, for instance, in noncommutative space \cite{bb}, with Coulomb-type scalar \cite{bb1} and vector potentials \cite{bb2}, under the influence of Coulomb-type and linear confining potential \cite{bb3}, with Cornell-type potential in fifth dimensional Minkowski space-time using the Kaluza-Klein theory \cite{EHR}.

Some Grand Unified Theories suggested that topological defects may have been produced during the phase transition in the early universe through a spontaneous symmetry breaking mechanism \cite{TWBK,AVV}. Various topological defects includes cosmic strings \cite{WAH,SZ2}, domain walls \cite{AVV,AVV2}, and global monopoles \cite{MBAV,DPB,TRPC,ERBM}. In condensed matter physics \cite{AVV3,TV,MOK,HK,TWBK2,CF22,CF33}, these linear defects are related to screw dislocations (or torsion) and disclinations (curvature) \cite{MOK,RAP}. A global monopole is a heavy object characterized by spherically symmetry and divergent mass. The gravitational field of a static global monopole was found by Barriola {\it et al.} \cite{MBAV} and are expected to be stable against spherical as well as polar perturbation. The effects of global monopole in quantum mechanical systems have been studied, for instance, in the non-relativistic limit, the harmonic oscillators \cite{CF44}, quantum scattering of charged or massive particles \cite{ALCO2,ERBM33,ERBM44}, solutions of the Klein-Gordon equation in the presence of a dyon, magnetic flux and scalar potential \cite{ALCO}. In the relativistic limit, studies on hydrogen and pionic atom \cite{ERFM2}, exact solutions of scalar bosons in the presence of Coulomb potential \cite{ABHA}, the Dirac and Klein–Gordon oscillators \cite{EAFB}, the generalized Klein-Gordon oscillator \cite{SZ}, and the Klein–Gordon oscillator with rainbow gravity \cite{SZ3}. In addition, global monopole has been studied, for instance, in scalar self-energy for a charged particle \cite{AAS,AAS2}, induced self-energy on a static scalar charged particle \cite{AAS3}, vacuum polarization for mass-less scalar fields \cite{AAS4}, vacuum polarization for mass-less spin-$1/2$ fields \cite{AAS5}, vacuum polarization effects in the presence of the Wu–Yang magnetic monopole \cite{AAS6}. Moreover, gravitational deflection of light by a rotating global monopole space-time have been investigated in Ref. \cite{AAS7} and more recently in Ref. \cite{AAS9}, and a charged global monopole in Ref. \cite{AAS8}. Other topological defects, such as cosmic strings which are one-dimensional linear defects have been studied in quantum system, for example, the Klein-Gordon oscillator in fifth-dimensional cosmic string space-time using the Kaluza-Klein theory \cite{JC}, the Klein-Gordon oscillator in a cosmic string space-time with an external magnetic field \cite{AB}, the Dirac field and oscillator in a spinning cosmic string \cite{MH5}, the Dirac oscillator under the influence of non-inertial effects in cosmic string space-time \cite{KB10}, the relativistic quantum dynamics of Klein–Gordon scalar fields subject to Cornell-type potential in a spinning cosmic-string space-time \cite{MH}, spin-zero bosons in an elastic medium with a screw dislocation \cite{SZ4}, the generalized DKP oscillator in a spinning cosmic string \cite{SZ5}, the generalized Klein-Gordon oscillator under a uniform magnetic field in a spinning cosmic string space-time \cite{cc1}, the modified Klein–Gordon oscillator under a scalar and electromagnetic potentials in rotating cosmic string space-time \cite{IJGMMP}, the spin-$0$ DKP equation and oscillator with a Cornell interaction in a cosmic-string space-time \cite{cc13}, the interaction of a Cornell-type non-minimal coupling with the scalar field under the topological defects \cite{cc14}, the influence of topological defects space–time with a spiral dislocation on spin-$0$ bosons field (via the DKP equation formalism) \cite{cc12}, and the generalized Klein-Gordon oscillator in fifth-dimensional cosmic string space-time without or with external potential in context of the Kaluza-Klein theory \cite{AHEP2,EPJC,SR}, relativistic vector bosons with non-minimal coupling in the spinning cosmic string space-time \cite{cc15}.

Our interest in magnetic monopole is due to the new reach made by the MoEDAL detector, an experiment for the search of magnetic monopole and other highly ionizing and long-lived particles at the CERN's Large Hadron Collider. MoEDAL is a largely passive detector that consists of three detection systems, namely, nuclear track detector, trapping detector array and time pix array \cite{BA}. In recent times, an analysis of 13-TeV pp collisions with the trapping detector during the 2015-2017 period provided the mass limits in the 1500–3750 GeV range of magnetic charges up to 5gD for monopole of spin-$0$, $\frac{1}{2}$ and $1$ in Ref. \cite{BA2}. MoEDAL also performed the first search for dyons, and based on a Drell-Yan production model, excluded dyons with a magnetic charge up to 5gD and electric charge up to 200e for the mass limits in the range 870-3120 GeV, and monopole with magnetic charge up to 5gD with the mass limits in the range 870–2040 GeV in Ref. \cite{BA3}. In this paper, we will study the relativistic quantum oscillator via the generalized Klein-Gordon oscillator formalism in the presence of the Aharonov-Bohm magnetic flux subject to Coulomb-type interaction potentials in a point-like global monopole space-time. We discuss the effects of topological defects, the magnetic flux, and interaction potentials on the energy profile and the wave function of these oscillator fields. Potential applications of this work include condensed matter systems, linear defects in elastic solids, impurities and vacancies in elastic continuous solids. This paper contains a theoretical analysis of relativistic quantum motions of spin-zero oscillator field by solving the generalized KG-oscillator in the point-like global monopole space–time subject to interaction potential.

This contribution is organized as follows: in {\tt section 2}, we discuss in details the generalized Klein-Gordon oscillator in the background space-time produced by a point-like global monopole, and then determine its solutions by choosing respectively, Coulomb-type function $f(r)=\frac{b}{r}$ without external potential ({\tt sub-section 2.1}), Cornell-type function $f(r)=\Big(a\,r+\frac{b}{r}\Big)$ without external potential ({\tt sub-section 2.2}), Coulomb-type function $f(r)=\frac{b}{r}$ with Coulomb-types scalar $S (r) (\propto \frac{1}{r})$ and vector $A_0 (\propto \frac{1}{r})$ potentials ({\tt sub-section 2.3}), Cornell-type function $f(r)=\Big(a\,r+\frac{b}{r}\Big)$ with Coulomb-types scalar $S (r) (\propto \frac{1}{r})$ and vector $A_0 (\propto \frac{1}{r})$ potentials ({\tt sub-section 2.4}); and finally conclusions in {\tt section 3}. Here, we have used the natural units $c=1=\hbar$.

\section{Generalized KG-Oscillator in a Global Monopole Space-time with Interaction Potential }
\label{sec: 1}

We study the quantum dynamics of oscillator fields in a space-time background produced by a point-like global monopole (PGM) without or with external potential. By solving the generalized Klein-Gordon oscillator analytically, we discuss the effects of the topological defects characterise by the parameter $\alpha$ of the space-time on the energy profile of these oscillator fields. Thereby, we begin this section by introducing the line element of a point-like global monopole which is a static and spherically symmetric metric in coordinates $(t, r, \phi, \theta)$ described by  \cite{CF44,ERBM44,ALCO,ERFM2,ABHA,EAFB,SZ,MBAV}
\begin{equation}
ds^2=-dt^2+\frac{dr^2}{\alpha^2}+r^2\,(d\theta^2+\sin^2 \theta\,d\phi^2),
\label{1}
\end{equation}
where $\alpha^2=\Big(1-8\,\pi\,\eta_0^2\Big)<1$ depends on the energy scale $\eta_0$. The parameter $\eta_0$ represents the dimensionless volumetric mass density of the point-like global monopole (PGM) defect. Here, the different coordinates are in the ranges $-\infty < t < +\infty$,\quad $0 \leq r < \infty$,\quad $0 \leq \theta \leq \frac{\pi}{2}$, and $0 \leq \phi < 2\,\pi$. This point-like global monopole space-time have some interesting features: $(i)$ it is not globally flat, and possesses a naked curvature singularity on the axis given by the Ricci scalar, $R=R^{\,\mu}_{\mu}=\frac{2\,(1-\alpha^2)}{r^2}$; $(ii)$ the area of a sphere of radius $r$ in this manifold is not $4\,\pi\,r^2$ but rather it is equal to $4\,\pi\,\alpha^2\,r^2$; $(iii)$ the surface $\theta=\frac{\pi}{2}$ presents the geometry of a cone with the deficit angle $\nabla\,\phi=8\,\pi^2\,\eta_0^2$; and $(iv)$ there is no Newtonian-like gravitational potential: $g_{tt}=-1$. Furthermore, in this topological defect space-time geometry the solid angle of a sphere of radius $r$ is $4\,\pi^2\,r^2\,\alpha^2$ which is smaller than $4\,\pi^2\,r^2$, and hence, there is a solid angle deficit $\nabla\,\Omega=32\,\pi^2\,\eta_0^2$. In condensed matter systems, this point-like global monopole space-time describes an effective metric produced in super-fluid $^{3}He$-by a monopole with the angle deficit $\alpha$. In that case, the topological defect has a negative mass \cite{GEV}. As this point-like global monopole geometry has no gravitational fields, some global effects of this geometry has been measured, for example, scattering cross section for mass-less bosonic \cite{POM}, and fermionic particles propagating on it \cite{HR}. Note that in the limit $\alpha \to 1$, one can obtain the Minkowski flat space line element in spherically symmetric.

The relativistic quantum dynamics of spin-zero scalar particles with an external potential $S(r)$ is described by the following wave equation \cite{ALCO,JC,AHEP2,EPJC,SR}
\begin{eqnarray}
\Bigg[-\frac{1}{\sqrt{-g}}\,D_{\mu}\,\Big\{\sqrt{-g}\,g^{\mu\nu}\,D_{\nu}\Big\}+\xi\,R+\Big(M+S(r)\Big) ^2\Bigg]\,\Psi=0,
\label{2}
\end{eqnarray}
where $D_{\mu} \equiv \Big(\partial_{\mu}-i\,e\,A_{\mu}\Big)$, $e$ is the electric charges, $A_{\mu}=(-A_0, \vec{A})$ is the electromagnetic four-vector potential, $g$ is the determinant of the metric tensor with $g^{\mu\nu}$ its inverse, $\xi$ is an arbitrary coupling constant with the background curvature, $R$ is the Ricci scalar or the scalar curvature, and $M$ is the rest mass of the scalar particle. Here we followed Refs. \cite{HGD,WG,WG2,AHEP2,EPJC,SR}, where it was suggested that a non-electromagnetic or static scalar potential $S (r)$ can introduce by modifying the rest mass of the scalar particle via transformation $M^2 \to \Big(M + S(r)\Big)^2$ in the wave equation. This new formalism has been first used in Ref. \cite{WG,WG2} to analyse the Dirac particle in the presence of a Coulomb and static scalar potentials proportional to the inverse of the radial distance, {\it i. e.}, $S(r) \propto \frac{1}{r}$. Later on, this new formalism has been studied in various space-times background in quantum systems by several researchers (Refs. \cite{AHEP2,EPJC,SR} and related references there in). 

In this contribution, we examine a spin-zero relativistic quantum oscillator (via the generalized Klein-Gordon oscillator) in the point-like global monopole (PGM) space-time background. We performed the replacement in the radial momentum vector $p_{\mu} \to \Big(p_{\mu}-i\,M\,\omega\,X_{\mu}\Big)$ and $p^{\,\dagger}_{\mu} \to \Big(p_{\mu}+i\,M\,\omega\,X_{\mu}\Big)$ \cite{AB,MH,JC,EHR,AHEP2,EPJC,SR} where, $\omega$ is the oscillation frequency and $X_{\mu}=f(r)\,\delta^{r}_{\mu}$ is a four-vector with $f(r) \neq r$ an arbitrary function. Note that if one choose $f(r)=r$ in the above replacement, {\it e. g.}, $p_{\mu} \to (p_{\mu}-i\,M\,\omega\,r\,\delta^{r}_{\mu})$ or $\vec{p} \to (\vec{p}-i\,M\,\omega\,\vec{r})$, then the quantum system is called the Klein-Gordon oscillator \cite{AB,MH,JC,EHR}. As we are focusing on the generalized Klein-Gordon oscillator, so we have replaced $r \to f(r)$, and, we can do replacement $\vec{p}^{\,\,2} \to \Big(\vec{p}-i\,M\,\omega\,f(r)\,\hat{r}\Big)\bullet \Big(\vec{p}+i\,M\,\omega\,f(r)\, \hat{r}\Big)$ in the wave equation. It is worth mentioning that this type of transformation where $r \to f(r)$ was first used in Ref. \cite{KBCF} to study the Dirac oscillator. So with the choice $X_{\mu}=(0, f(r), 0, 0)$, the invariant of the Lorentz transformation of particle (or the active invariant) is broken. Indeed, this coupling preserves the invariant of the Lorentz transformation of observers as the behaviour of a genuine background field.  

Therefore, the generalized Klein-Gordon oscillator from Eq. (\ref{2}) becomes
\begin{eqnarray}
\Bigg[-\frac{1}{\sqrt{-g}}\,\Big(D_{\mu}+M\,\omega\,X_{\mu}\Big)\,\Big\{\sqrt{-g}\,g^{\mu\nu}\,\Big(D_{\nu}-M\,\omega\,X_{\nu}\Big)\Big\}+\xi\,R+\Big(M+S(r)\Big)^2\Bigg]\,\Psi=0,
\label{3}
\end{eqnarray}

By the method of separation of variables, one can always write the total wave function $\Psi (t, r, \theta, \phi)$ in terms of different variables. Suppose, we choose a possible wave function in terms of a radial wave function $\psi (r)$ as :
\begin{equation}
\Psi(t, r, \theta, \phi)=e^{-i\,E\,t}\,Y_{l,m} (\theta, \phi)\,\psi (r),
\label{4}
\end{equation}
where $E$ is the energy of the scalar particles, $Y_{l,m} (\theta, \phi)$ is the spherical harmonics, and $l, m$ are respectively the angular momentum and magnetic moment quantum numbers.

Also, we have chosen the three-vector electromagnetic potential $\vec{A}$ given by Refs. \cite{ALCO,ABHA}
\begin{equation}
A_{r}=0=A_{\theta},\quad A_{\phi}=\frac{\Phi_B}{2\,\pi\,r\,\sin \theta},\quad \Phi_B=\Phi\,\Phi_0,\quad \Phi_0=2\,\pi\,e^{-1},
\label{5}
\end{equation}
where $\Phi_B=const$ is the Aharonov-Bohm magnetic flux, $\Phi_0$ is the quantum of magnetic flux, and $\Phi$ is the amount of magnetic flux which is a positive integer. Note that the Aharonov–Bohm effect \cite{YA,MP} is a quantum mechanical phenomena and has been investigated in different branches of physics including bound-states of massive fermions \cite{VRK2}, and in the context of Kaluza–Klein theory \cite{JC,EHR,AHEP2,EPJC,SR} etc.. 

Thereby, in the space-time background (\ref{1}) and using Eqs. (\ref{4})--(\ref{5}) into the Eq. (\ref{3}), we obtain the following differential equation:
\begin{eqnarray}
&&\frac{1}{\psi}\left[\alpha^2\,\left\{\left(\frac{\partial}{\partial r} + M\,\omega\,f\right) \left(r^2\,\frac{\partial \psi}{\partial r}-M\,\omega\,r^2\,f\,\psi\right)\right\}\right]+\frac{1}{\psi}\left[\left\{-\frac{2\,\xi\,(1-\alpha^2)}{r^2}+\Big(E-e\,A_0\Big)^2-\Big(M+S(r)\Big)^2\right\}\,\psi\right]\nonumber\\
&+&\frac{1}{r^2\,Y_{l,m}}\,\left[\frac{1}{\sin\theta}\,\frac{\partial}{\partial \theta}\,\left(\sin\theta\, \frac{\partial Y_{l,m}}{\partial \theta}\right)+\frac{1}{\sin^2\theta}\,\left(\frac{\partial}{\partial \phi}-i\,\Phi\right)^2\,Y_{l,m} \right]=0.
\label{6}
\end{eqnarray}
The effective total momentum and the $z$-component of the angular momentum operators are follows:
\begin{eqnarray}
L^2_{eff}=\left[-\frac{1}{\sin\theta}\,\frac{\partial}{\partial \theta}\,\left(\sin \theta\,\frac{\partial} {\partial \theta}\right)-\frac{1}{\sin^2\theta}\,\left(\frac{\partial}{\partial \phi}-i\,\Phi\right)^2\right] ,\quad L^{eff}_{z}=-i\,\left(\frac{\partial}{\partial \phi}-i\,\Phi\right).
\label{7}
\end{eqnarray}
The eigenvalues of the various operator terms involve in Eq. (\ref{7}) are as follows \cite{ALCO,ABHA} 
\begin{eqnarray}
&&L_{z}\,Y_{l,m} (\theta, \phi)=m\,Y_{l,m} (\theta, \phi)\quad,\quad L^{eff}_{z}\,Y_{l,m} (\theta, \phi)=(m-\Phi)\,Y_{l,m} (\theta, \phi),\nonumber\\
&&L^2\,Y_{l,m} (\theta, \phi)=\lambda\,Y_{l,m} (\theta, \phi)\quad,\quad L^2_{eff}\,Y_{l,m} (\theta, \phi)=\lambda'\,Y_{l,m} (\theta, \phi)\nonumber\\
&&\lambda=l\,(l+1)\quad,\quad \lambda'=l'\,(l'+1)\quad,\quad l'=l-\Phi.
\label{8}
\end{eqnarray}
 
Thereby, using the above operators in Eq. (\ref{6}), the radial wave equation for the generalized Klein-Gordon oscillator becomes:
\begin{eqnarray}
&&\psi''(r)+\frac{2}{r}\,\psi'(r)-\Big[2\,M\,\omega\,\Big(\frac{f}{r}+\frac{f'}{2}\Big)+M^2\,\omega^2\,f^2 (r)\Big]\,\psi (r)\nonumber\\ &+&\frac{1}{\alpha^2}\,\Bigg[-\frac{2\,\xi\,(1-\alpha^2)+l'\,(l'+1)}{r^2}+\Big(E-e\,A_0\Big)^2-\Big(M+S ( r)\Big)^2\Bigg]\,\psi (r)=0.\quad\quad
\label{9}
\end{eqnarray}
 
To study the generalized Klein-Gordon oscillator in a point-like global monopole space-time, we have chosen two types of  function $f(r)$, namely, a Coulomb-type function ($f(r) \propto \frac{1}{r}$), and a Cornell-type potential form function (linear plus Coulomb-type function) and obtain the bound-states energy eigenvalues and the wave function of these oscillator fields.

\subsection{\bf Coulomb-Type Potential Form Function $f(r)=\frac{b}{r}$ Without External Potential $A_0=0=S$.}

In this section, we study the generalized Klein-Gordon oscillator by choosing a function $f(r)$ proportional to the inverse of the radial distance, {\it i. e.,} $f (r) \propto \frac{1}{r}\Rightarrow f(r)=\frac{b}{r}$ where, $b>0$ is a constant in the point-like global monopole (PGM) space-time subject to zero scalar and vector potential, $A_0=0=S$. Several authors have been used this type of function to study the generalized KG- and/or the Dirac oscillator in quantum systems Refs. \cite{AHEP2,EPJC,SZ2,SZ,IJGMMP}. We discuss the effects of the topological defects of the space-time as well as the above function $f (r)$ on the energy profiles of these oscillator fields.

Thereby, substituting the function $f(r)=\frac{b}{r}$ and considering $A_0=0=S$ into the Eq. (\ref{9}), the radial wave equation becomes
\begin{equation}
\psi''(r)+\frac{2}{r}\,\psi'(r)+\Big(\zeta^2-\frac{\sigma^2}{r^2}\Big)\,\psi (r)=0,
\label{11}
\end{equation}
where
\begin{equation}
\zeta=\frac{\sqrt{E^2-M^2}}{\alpha},\quad \sigma=\sqrt{\frac{2\,\xi\,(1-\alpha^2)+l'\,(l'+1)}{\alpha^2}+M\,\omega\,b\,+M^2\,\omega^2\,b^2}.
\label{12}
\end{equation}
Now, we perform a transformation via $\psi (r)=\frac{U (r)}{\sqrt{r}}$ into the Eq. (\ref{11}), we have
\begin{equation}
r^2\,U'' (r)+r\,U' (r)+\Big(\zeta^2\,r^2-\tau^2\Big)\,U (r)=0,
\label{13}
\end{equation}
where $\tau=\sqrt{\sigma^2+\frac{1}{4}}$. 

Equation (\ref{13}) is the well-known Bessel's differential equation \cite{GBA}. Since $\tau$ is always positive, the general solution to the Bessel equation (\ref{13}) is in the form: $U(r)=C_1\,J_{|\tau|} (\zeta\,r)+C_2\,Y_{|\tau|} (\zeta\,r)$, where $J_{|\tau|} (\zeta\,r)$ and $Y_{|\tau|} (\zeta\,r)$ are the Bessel function of first kind and second kind \cite{GBA}, respectively. The Bessel function of second kind $Y_{|\tau|} (\zeta\,r)$ diverges at the origin; so, we must take $C_2=0$ in the general solution, since we are mainly interested in the well-behaved solution. Thus, the regular solution to the Eq. (\ref{13}) at the origin is given by
\begin{equation}
U (r)=C_1\,J_{|\tau|} (\zeta\,r).
\label{14}
\end{equation}
Let us restrict the motions of scalar fields to the region where a hard-wall potential is present. This kind of confinement is described by the boundary condition: $U (r_0)=0$, which means that the wave function $\psi (r)$ vanishes at a fixed radius $r=r_0$; that is, this boundary condition corresponds to the scalar field subject to a hard-wall confining potential. This type confining potential has been studied in quantum system, for instance, on scalar particles under non-inertial effects \cite{mm6,mm3} (and relatd references there in). Let us consider a particular case where $\zeta\,r_0 > > 1$. In this particular case, we can write (\ref{14}) in the following form
\begin{equation}
J_{|\tau|} (\zeta\,r_0) \propto \cos\,\Big(\zeta\,r_0-\frac{\tau\,\pi}{2}-\frac{\pi}{4}\Big).
\label{15}
\end{equation}
Thereby, substituting Eq. (\ref{15}) into the Eq. (\ref{14}) and using the boundary condition $U (r_0)=0$, one can find the relativistic energy profile of these oscillator fields as follows:
\begin{equation}
E_{n,l}=\pm\,\sqrt{M^2+\Bigg\{n+\frac{1}{2}\,\Bigg(\sqrt{\frac{1}{4}+\frac{2\xi\,(1-\alpha^2)+l'\,(l'+1)}{\alpha^2}+M\,\omega\,b+M^2\,\omega^2\,b^2}+\frac{3}{2}\Bigg)\Bigg\}^2\,\frac{\alpha^2\,\pi^2}{r^2_0}},
\label{16}
\end{equation}
where $l'=(l-\Phi)$ and $n=0,1,2,3,...$.

\begin{figure}
\begin{subfigure}[b]{0.42\textwidth}
\includegraphics[width=3.2in,height=2.1in]{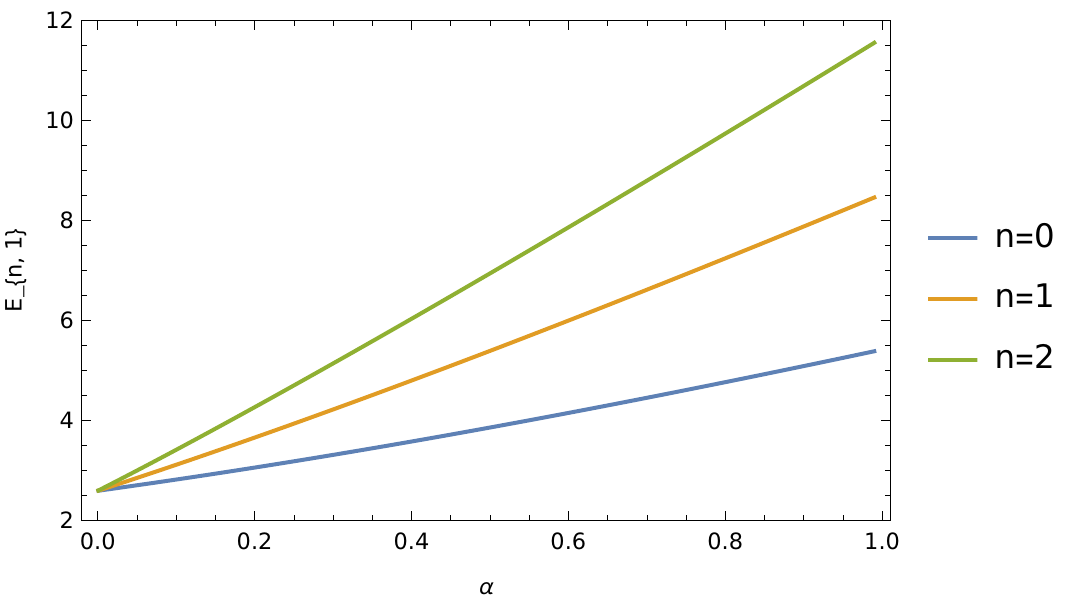}
\caption{$l=1=\omega, \Phi=1/4$}
\label{fig: (a)}
\end{subfigure}
\hfill
\begin{subfigure}[b]{0.42\textwidth}
\includegraphics[width=3.2in,height=2.1in]{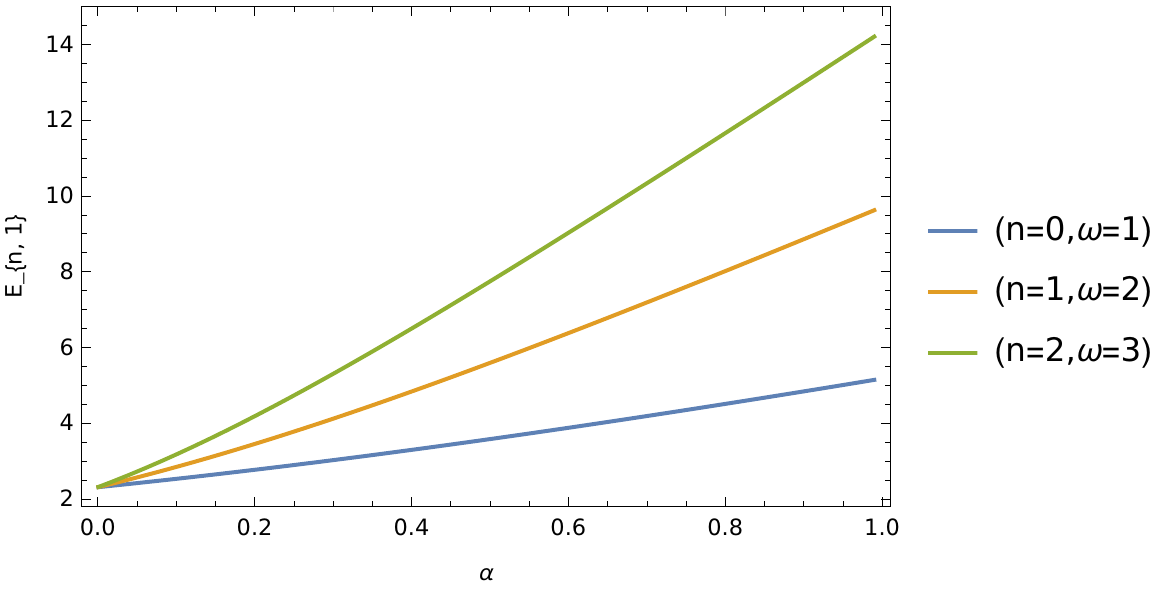}
\caption{$l=1$, $\Phi=1/2$}
\label{fig: (b)}
\end{subfigure}
\hfill\\
\begin{subfigure}[b]{0.4\textwidth}
\includegraphics[width=3.2in,height=2.1in]{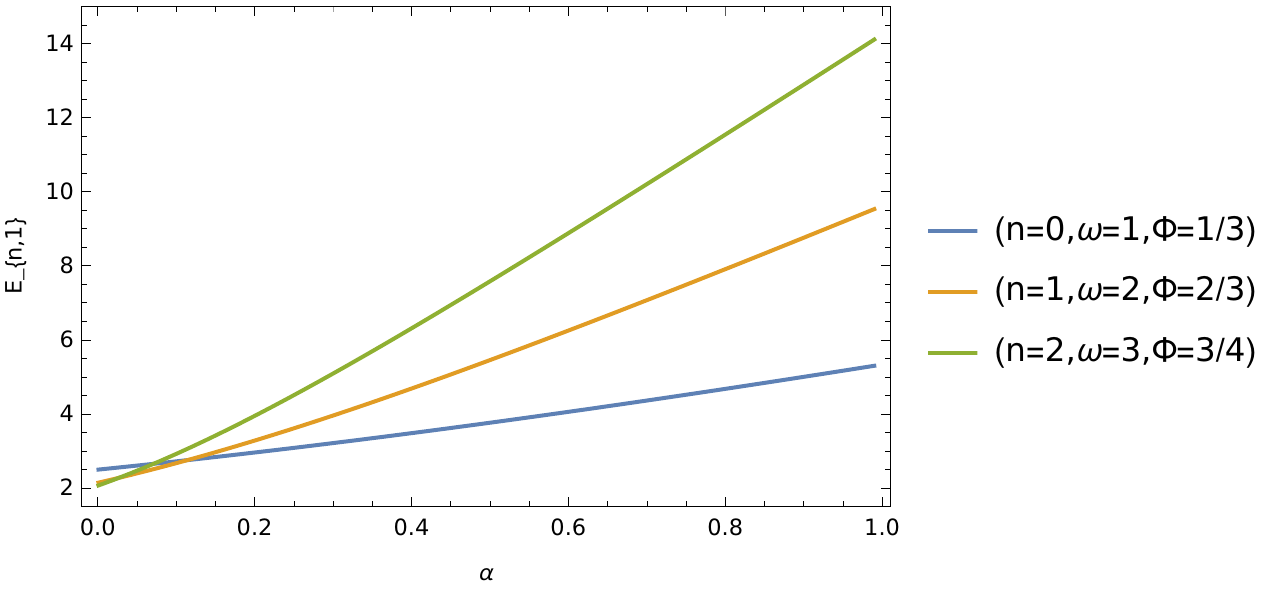}
\caption{$l=1$}
\label{fig: (c)}
\end{subfigure}
\hfill
\begin{subfigure}[b]{0.42\textwidth}
\includegraphics[width=3.2in,height=2.1in]{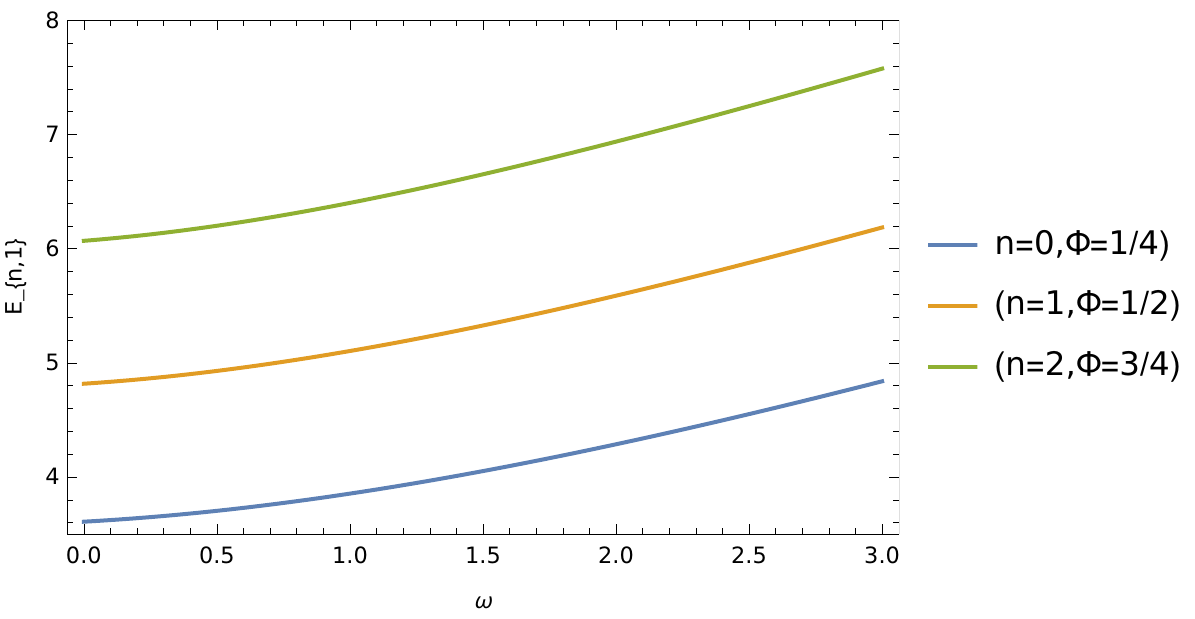}
\caption{$l=1$, $\alpha=1/2$}
\label{fig: (d)}
\end{subfigure}
\caption{Eigenvalues $E_{n,l}$ keeping fixed $M=1=r_0=b$, $\xi=1/2$ [unit of different parameters are chosen in natural units $(c=1=\hbar)$].}
\label{fig: 1}
\end{figure}

Thus we can see that the background space-time produced by a point-like global monopole modified the relativistic energy eigenvalues of these oscillator fields through the presence of the topological defect parameter $\alpha^2$ subject to a hard-wall confining potential. In addition, we see that the energy eigenvalues $E_{n,l}$ depends on the geometric quantum phase $\Phi_B$, and this dependence of the eigenvalue on the quantum phase gives us the gravitational analogue to the Aharonov-Bohm effect \cite{YA,MP}. We have plotted few graphs showing the effects of different parameters on the energy spectrum of these oscillator fields (fig. 1).

\subsection{\bf Cornell-Type Potential Form Function $f(r)=\Big(a\,r+\frac{b}{r}\Big)$ Without External Potential $A_0=0=S$.}

In this section, we study the generalized Klein-Gordon oscillator in the point-like global monopole space-time by choosing the Cornell-type potential form function $f(r)=\Big(a\,r+ \frac{b}{r}\Big)$ \cite{KBCF} with zero scalar and vector potential, $A_0=0=S$. This type of function has been used for the studies of the generalized Klein-Gordon oscillator in Refs. \cite{AHEP2,EPJC,SR,SZ2,SZ,cc1,IJGMMP}, and the generalized Dirac oscillator in Refs. \cite{LFD} in quantum systems. We solve the generalized KG-oscillator analytically and discuss the effects of topological defects as well as this type of function on the energy profile of these oscillator fields.

Thereby, substituting the Cornell-type function $f(r)=\left(a\,r+\frac{b}{r}\right)$ and considering $A_0=0=S$ into the Eq. (\ref{9}), we have obtained the following radial wave equation:
\begin{equation}
\psi''(r)+\frac{2}{r}\,\psi'(r)+\Big(\Lambda-\frac{j^2}{r^2}-M^2\,\omega^2\,a^2\,r^2\Big)\,\psi(r)=0,
\label{aa1}
\end{equation}
where we have defined different parameters
\begin{eqnarray}
\Lambda=\frac{E^2-M^2-3\,M\,\omega\,a\,\alpha^2-2\,a\,b\,M^2\,\omega^2\,\alpha^2}{\alpha^2},\quad
j=\sqrt{\frac{2\,\xi\,(1-\alpha^2)+l'\,(l'+1)}{\alpha^2}+M\,\omega\,b+M^2\,\omega^2\,b^2}.
\label{aa2}
\end{eqnarray}
Transforming the above equation (\ref{aa2}) via $\psi (r)=\frac{U (r)}{r^{3/2}}$, we have 
\begin{equation}
U''(r)-\frac{1}{r}\,U'(r)+\Bigg[\Lambda-\frac{(j^2-\frac{3}{4})}{r^2}-M^2\,\omega^2\,a^2\,r^2\Bigg]\,U (r)=0.
\label{aa3}
\end{equation}
Introducing a new variables via $s=M\,\omega\,a\,r^2$ into the above Eq. (\ref{aa3}), we have obtained the following second order differential equation:
\begin{equation}
U''(s)+\left(\frac{1-4\,\mu^2}{4\,s^2}\right)\,U(s)+\frac{\nu}{s}\,U(s)-\frac{1}{4}\,U(s)=0,
\label{aa4}
\end{equation}
where different parameters are defined as
\begin{equation}
2\,\mu=\sqrt{j^2+\frac{1}{4}}\quad,\quad \nu=\frac{\Lambda}{4\,M\,\omega\,a}.
\label{aa5}
\end{equation}

Equation (\ref{aa4}) is the Whittaker differential equation \cite{KDM} and $U(s)$ is the Whittaker function which can be written in terms of the confluent hypergeometric function ${}_1 F_{1}(s)$ as
\begin{equation}
U (s)=s^{\frac{1}{2}+\mu}\,e^{-\frac{s}{2}}\,{}_1 F_{1}\left(\mu-\nu +\frac{1}{2},2\,\mu+1;s\right).
\label{aa6}
\end{equation}
In order to obtain the bound-states solutions, it is necessary that the confluent hypergeometric function ${}_1 F_{1}\left(\mu-\nu +\frac{1}{2},2\,\mu+1 ; s\right)$ must be a power series expansion of $s$ with degree $n$, and the quantity $\Big(\mu-\nu +\frac{1}{2}\Big)$ should be a negative integer, that is, $\Big(\mu-\nu+\frac{1}{2}\Big)=-n$, where $n=0,1,2,..$. After simplification of this condition, we have obtained the following eigenvalues expression for the quantum system
\begin{eqnarray}
E_{n,l}=\pm\,\sqrt{\begin{aligned}
&M^2\,\Big(1+2\,a\,b\,\omega^2\,\alpha^2\Big)+2\,M\,\omega\,\alpha^2\\ &\times\left(2\,n+\sqrt{\frac{1}{4}+\frac{2\,\xi\,(1-\alpha^2)+l'\,(l'+1)}{\alpha^2}+M\,\omega\,b+M^2\,\omega^2\,b^2}+\frac{5}{2}\right)
\end{aligned}
}.\quad\quad
\label{aa7}
\end{eqnarray}

\begin{figure}
\begin{subfigure}[b]{0.42\textwidth}
\includegraphics[width=3.2in,height=2.2in]{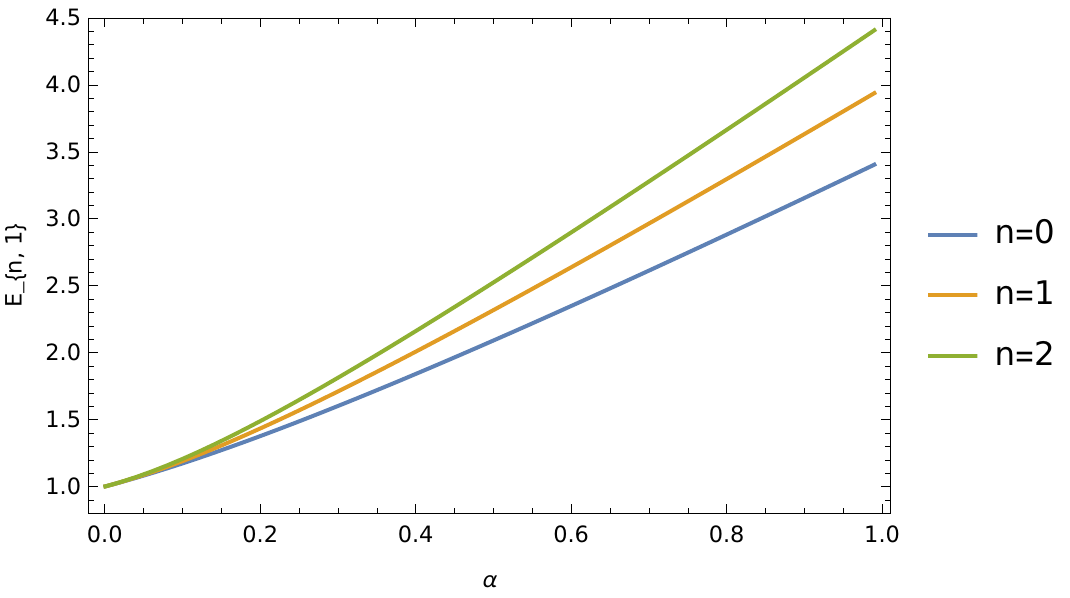}
\caption{$l=1=\omega, \Phi=1/4$}
\label{fig: 2 (a)}
\end{subfigure}
\hfill
\begin{subfigure}[b]{0.42\textwidth}
\includegraphics[width=3.2in,height=2.2in]{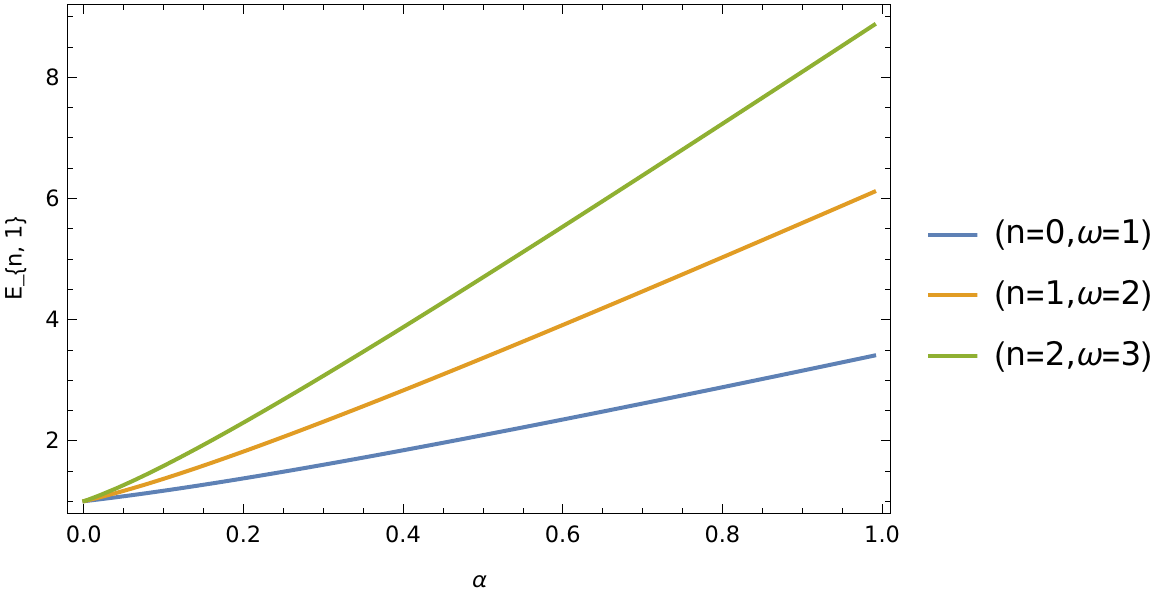}
\caption{$l=1, \Phi=1/4$}
\label{fig: 2 (b)}
\end{subfigure}
\hfill\\
\begin{subfigure}[b]{0.42\textwidth}
\includegraphics[width=3.2in,height=2.2in]{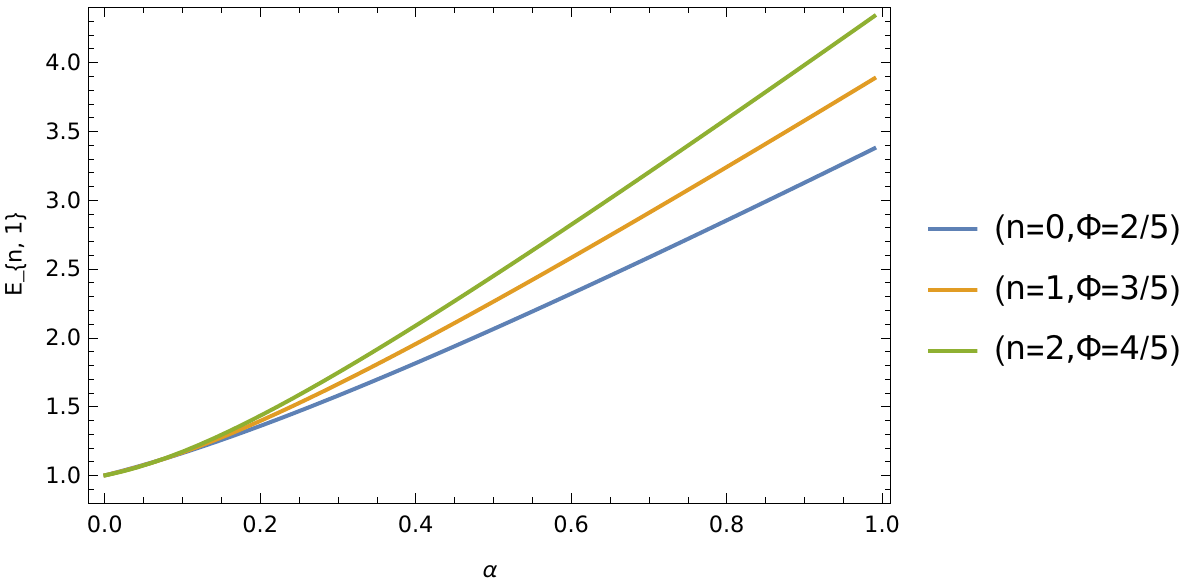}
\caption{$l=1=\omega$}
\label{fig: 2 (c)}
\end{subfigure}
\hfill
\begin{subfigure}[b]{0.42\textwidth}
\includegraphics[width=3.2in,height=2.2in]{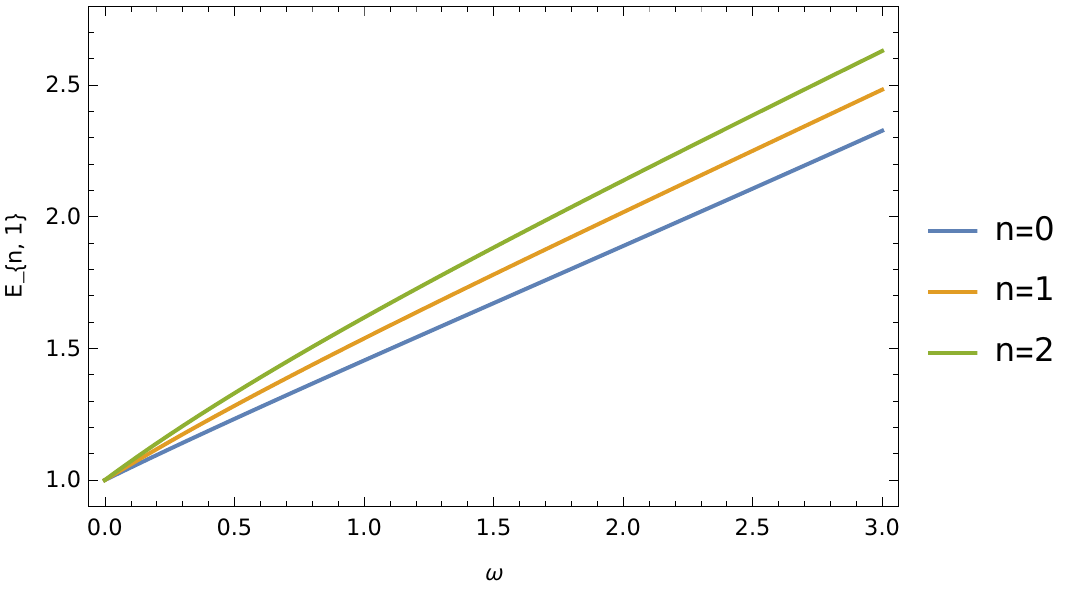}
\caption{$l=1, \alpha=1/4, \Phi=1/2$}
\label{fig: 2 (d)}
\end{subfigure}
\hfill\\
\begin{subfigure}[b]{0.42\textwidth}
\includegraphics[width=3.2in,height=2.2in]{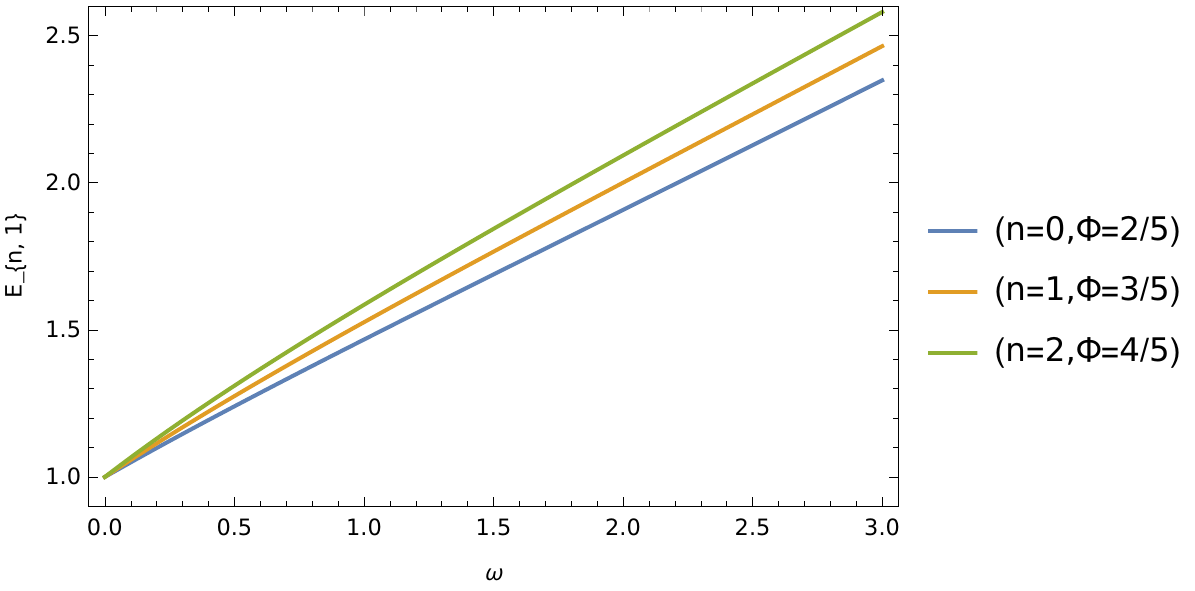}
\caption{$l=1, \alpha=1/4$}
\label{fig: 2 (e)}
\end{subfigure}
\hfill
\begin{subfigure}[b]{0.42\textwidth}
\includegraphics[width=3.2in,height=2.2in]{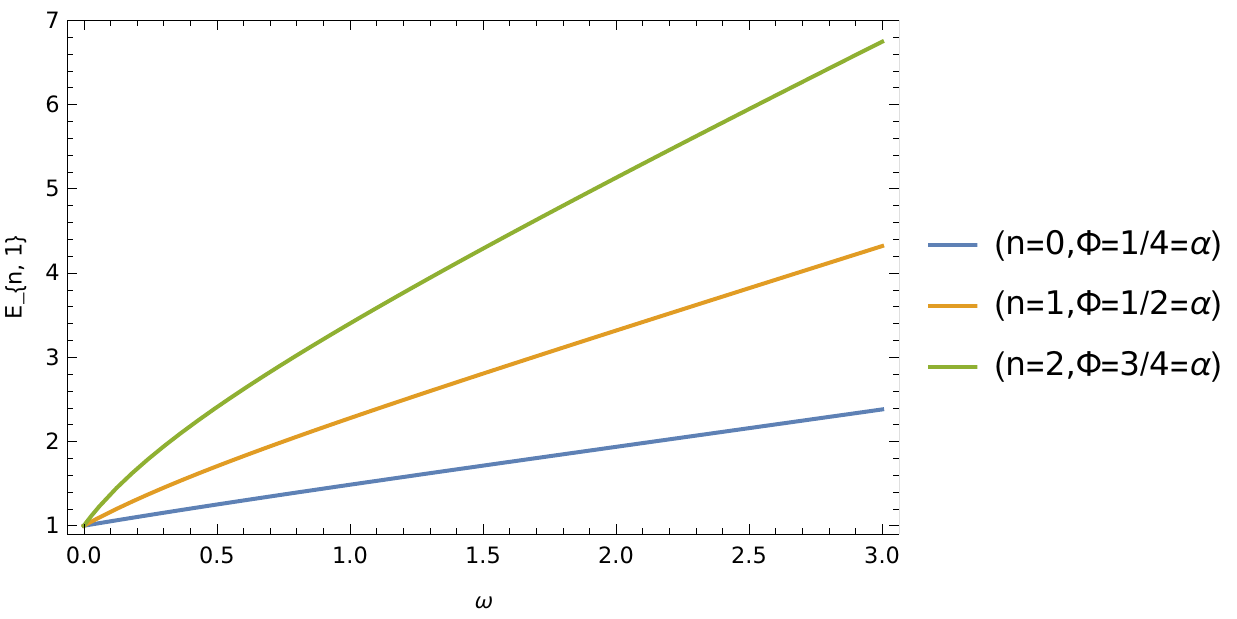}
\caption{$l=1$}
\label{fig: 2 (f)}
\end{subfigure}
\caption{Eigenvalues $E_{n,1}$ keeping fixed $M=1=b=a$, $\xi=1/2$ [unit of different parameters are chosen in natural units $(c=1=\hbar)$].}
\label{fig: 2}
\end{figure}

The normalized radial wave functions are given by
\begin{equation}
\psi_{n,l} (s)=D_{n,l}\,(M\,\omega\,a)^{3/4}\,s^{\mu-\frac{1}{4}}\,e^{-\frac{s}{2}}\,{}_1 F_{1}\left(\mu-\nu +\frac{1}{2},2\,\mu+1 ; s\right),
\label{aa8}
\end{equation}

\begin{figure}
\begin{subfigure}[b]{0.42\textwidth}
\includegraphics[width=3.0in,height=2.0in]{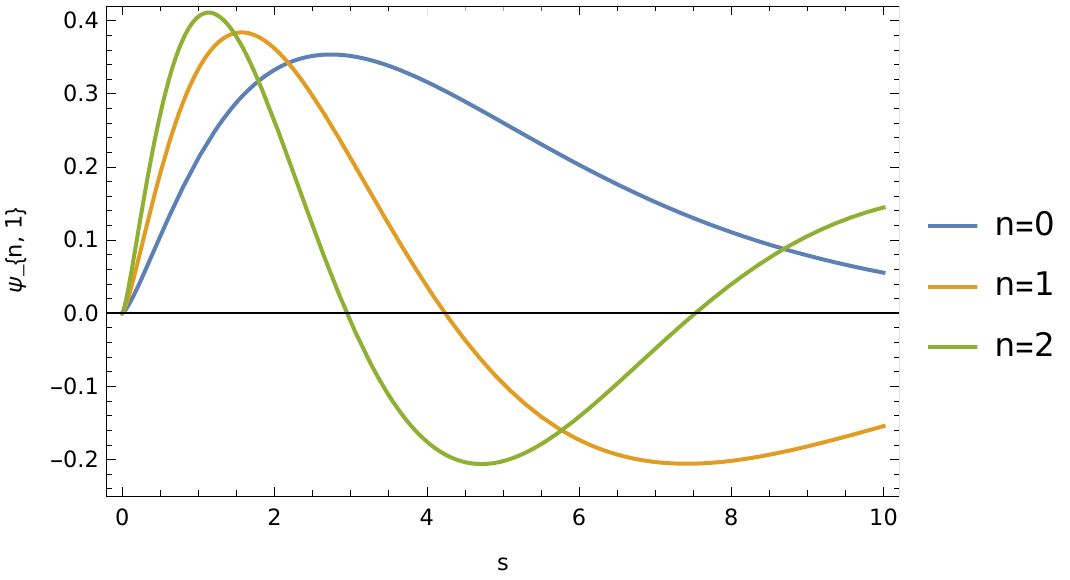}
\caption{$l=1=\omega, \Phi=1/4, \alpha=1/2$}
\label{fig: 3 (a)}
\end{subfigure}
\hfill
\begin{subfigure}[b]{0.42\textwidth}
\includegraphics[width=3.0in,height=2.0in]{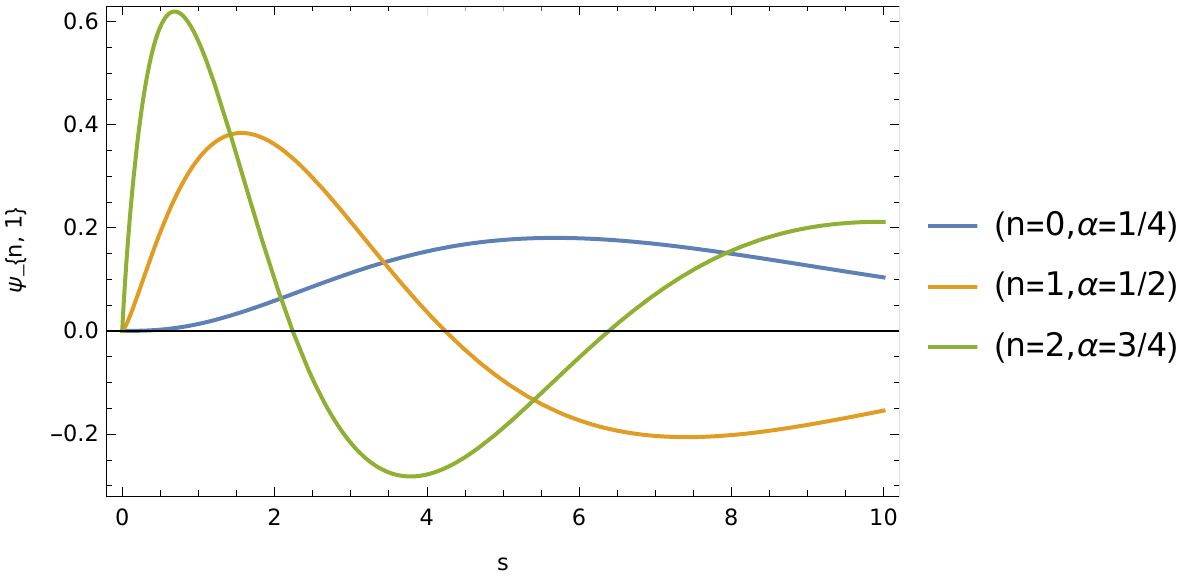}
\caption{$l=1=\omega, \Phi=1/4$}
\label{fig: 3 (b)}
\end{subfigure}
\hfill\\
\begin{subfigure}[b]{0.45\textwidth}
\includegraphics[width=3.2in,height=2.0in]{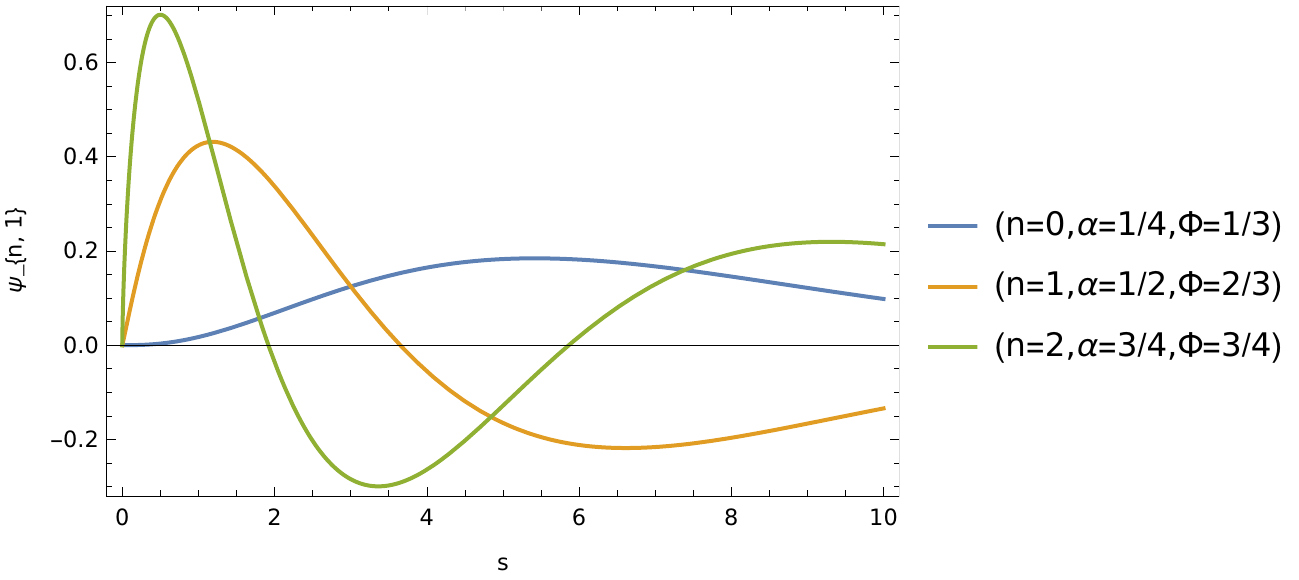}
\caption{$l=1=\omega$}
\label{fig: 3 (c)}
\end{subfigure}
\hfill
\begin{subfigure}[b]{0.42\textwidth}
\includegraphics[width=3.2in,height=2.0in]{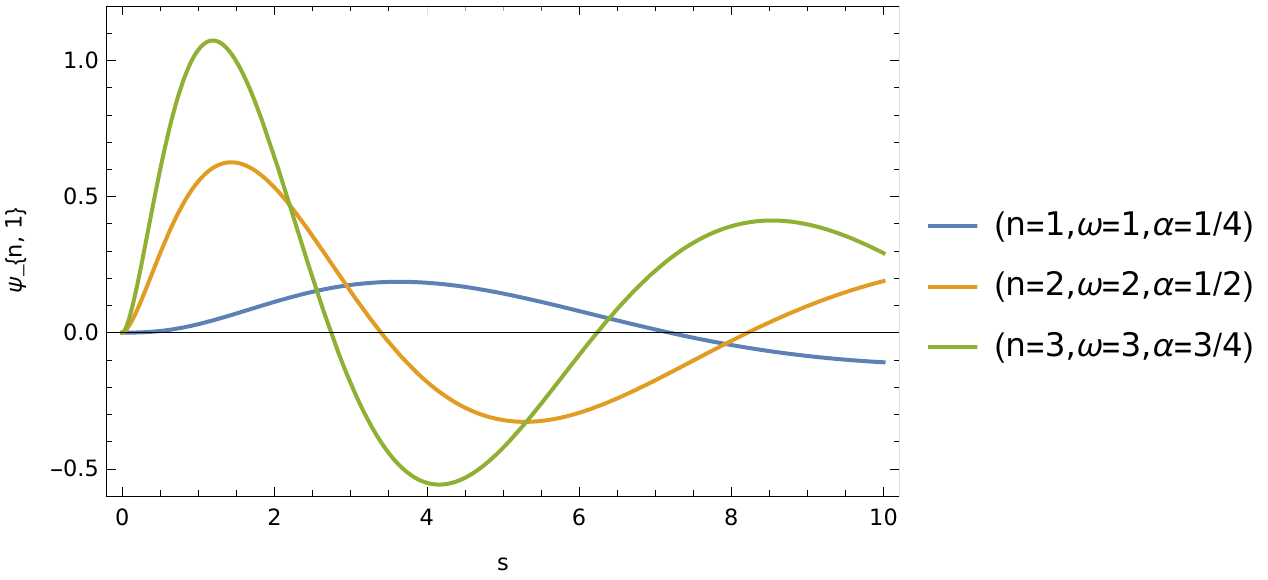}
\caption{$l=1, \Phi=1/4$}
\label{fig: 3 (d)}
\end{subfigure}
\hfill\\
\begin{subfigure}[b]{0.45\textwidth}
\includegraphics[width=3.2in,height=2.0in]{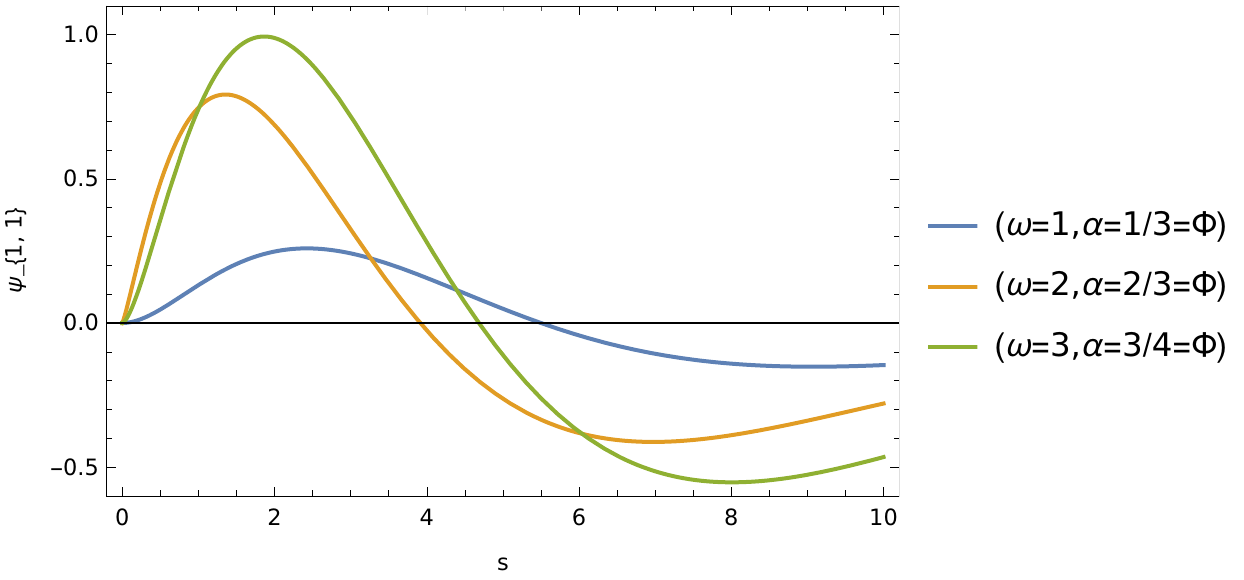}
\caption{$n=1=l$}
\label{fig: 3 (e)}
\end{subfigure}
\hfill
\begin{subfigure}[b]{0.42\textwidth}
\includegraphics[width=3.0in,height=2.0in]{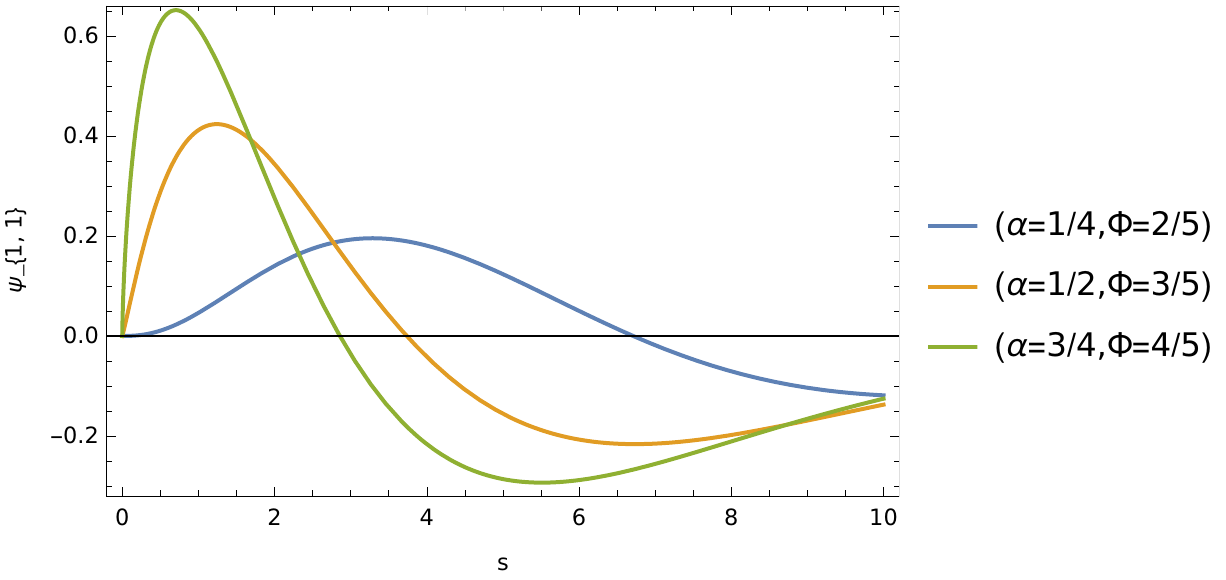}
\caption{$n=1=l=\omega$}
\label{fig: 3 (f)}
\end{subfigure}
\caption{Radial wave function $\psi_{n,l}$ keeping fixed $M=1=b=a$, $\xi=1/2$ [unit of different parameters are chosen in natural units $(c=1=\hbar)$].}
\label{fig: 3}
\end{figure}

where $D_{n,l}$ is a constant which can be determined by the normalization condition for the radial wave function
\begin{equation}
\frac{1}{\alpha}\,\int^{\infty}_{0}\,r^2\,dr\,|\psi (r)|^2=1.
\label{aa9}
\end{equation}
To solve the integrals of the radial wave function, we can write the confluent hypergeometric function in terms of the associated Laguerre polynomials by the relation \cite{GBA}
\begin{equation}
{}_1 F_{1}\left(-n,2\,\mu+1; x\right)=\frac{n!\,(2\,\mu)!}{(n+2\,\mu)!}\,L^{(2\,\mu)}_{n} (x).
\label{aa10}
\end{equation}
Then, taking into account $s=M\,\omega\,r^2$, and with the help of \cite{AP} to solve the integrals, the normalization constant is given by
\begin{equation}
D_{n,l}=\frac{1}{(2\,\mu)!}\,\sqrt{\frac{2\,\alpha\,(n+2\,\mu)!}{n!}}=\frac{1}{\left(\sqrt{j^2+\frac{1}{4}}\right)!}\,\sqrt{\frac{2\,\alpha\,\left(n+\sqrt{j^2+\frac{1}{4}}\right)!}{n!}}.
\label{aa11}
\end{equation}
Equation (\ref{aa7}) is the relativistic energy spectrum that stems from the interaction of the scalar field with the generalized Klein-Gordon oscillator in a point-like global monopole space-time background. We see that the topological defects associated with the scalar curvature of the space-time, and the Cornell-type function $f(r)=\Big(a\,r+\frac{b}{r}\Big)$ modified the energy profiles of these oscillator fields. Furthermore, the energy eigenvalues $E_{n,l}$ depends on the geometric quantum phase $\Phi_B$ which gives us the gravitational analogue to the Aharonov-Bohm effect \cite{YA,MP}. We have plotted graphs showing the effects of different parameters on the energy spectrum (fig. 2), and the wave function (fig. 3) of these oscillator fields.

In the below analysis, we insert an external potential through a minimal substitution to study the generalized Klein-Gordon oscillator in the point-like global monopole space-time with the function $f(r)$ considered in the previous two analysis. We have chosen the first component of the electromagnetic four-vector potential $A_0$ and a static scalar potential proportional to the inverse of the radial distance \cite{bb1,bb2,bb3}, {\it i. e.}, 
\begin{equation}
A_0 \propto \frac{1}{r}\Rightarrow A_0=\frac{\kappa}{r}=\pm\,\frac{|\kappa|}{r}\quad,\quad S \propto \frac{1}{r}\Rightarrow S=\frac{\eta}{r},
\label{potential}
\end{equation}
where $\kappa, \eta$ characterize the potential parameters. This type of potential have widely been used in different branches of physics, for instance, $1$-dimensional systems \cite{aa4,aa5}, topological defects in solids \cite{aa20}, quark-antiquark interaction \cite{aa30}, propagation of gravitational waves \cite{aa21}, quark models \cite{aa22}, and in the relativistic quantum systems \cite{mm3,IJGMMP}. Here we followed Refs. \cite{HGD,WG,WG2} to introduce the scalar potential in the wave equation by modifying the mass term via transformation $M \to M + S(t, r)$, where $S(t, r)$ is the scalar potential.

\subsection{\bf Coulomb-Type Function $f (r)=\frac{b}{r}$ With Coulomb-Type Vector $A_0=\pm\,\frac{|\kappa|}{r}$ and Scalar $S=\frac{\eta}{r}$ Potentials.}

In this section, we study the generalized Klein-Gordon oscillator subject to a vector and scalar potentials of Coulomb-types (\ref{potential}) with a Coulomb-type function $f(r)=\frac{b}{r}$ in the presence of an Aharonov-Bohm magnetic flux field in the space-time background produced by a point-like global monopole. 

Thereby, substituting the function $f(r)=\frac{b}{r}$ and Coulomb-types scalar and vector potential into the Eq. (\ref{9}), we have obtained the following radial wave equation: 
\begin{equation}
\psi''(r)+\frac{2}{r}\,\psi' (r)+\left(-\chi^2-\frac{\beta^2}{r^2}-\frac{2\,\gamma}{r}\right)\,\psi (r)=0,
\label{cc1}
\end{equation}
where we have defined different parameters
\begin{eqnarray}
\beta&=&\sqrt{\frac{2\,\xi\,(1-\alpha^2)+l'\,(l'+1)+\eta^2-e^2\,\kappa^2}{\alpha^2}+M\,\omega\,b+M^2\,\omega^2\,b^2},\quad \gamma=\frac{(M\,\eta-e\,E\,\kappa)}{\alpha^2},\nonumber\\
\chi&=&\frac{\sqrt{M^2-E^2}}{\alpha}.
\end{eqnarray}
Following the radial wave transformation via $\psi (r)=\frac{U (r)}{\sqrt{r}}$ into the Eq. (\ref{cc1}), we have
\begin{equation}
U'' (r)+\frac{1}{r}\,U' (r)+\left(-\chi^2-\frac{\beta^2}{r^2}-\frac{2\,\gamma}{r} \right)\,U (r)=0.
\label{cc2}
\end{equation}
We do another transformation on the radial coordinate via $\rho=2\,\chi\,r$ into the Eq. (\ref{cc2}), we have
\begin{equation}
U'' (\rho)+\frac{1}{\rho}\,U' (\rho)+\left(-\frac{\beta^2}{\rho^2}-\frac{\gamma}{\chi\,\rho}-\frac{1}{4} \right)\,U (\rho)=0.
\label{cc3}
\end{equation}
Now, we impose requirement of the wave function that the radial wave function $R(\rho)$  must be well-behaved at the origin, since it is a singular point of the Eq. (\ref{cc3}). In this case, for $lim_{\rho \to 0} U (\rho)=0$, the solution is $U (\rho) \sim \rho^{\beta}$. Furthermore, for $lim_{\rho \to \infty} U (\rho) =0$, the solution is $U (\rho) \sim e^{-\frac{\rho}{2}}$. Thus, a possible solution to the Eq. (\ref{cc3}) is given by
\begin{equation}
U (\rho)=\rho^{\beta}\,e^{-\frac{\rho}{2}}\,F (\rho), 
\label{cc4}
\end{equation}
where $F (\rho)$ is an unknown function. Substituting this solution (\ref{cc4}) into the Eq. (\ref{cc3}), we have
\begin{equation}
\rho\,F''(\rho)+(2\,\beta+1-\rho)\,F'(\rho)+\left(-\beta-\frac{\gamma}{\chi}-\frac{1}{2}\right)\,F (\rho)=0.
\label{cc5}
\end{equation}
Equation (\ref{cc5}) is the confluent hypergeometric second order differential equation and the function $F (\rho)= {}_1 F_{1} (\beta+\frac{\gamma}{\chi}+\frac{1}{2}, 2\,\beta+1 ; \rho)$ is called the confluent hypergeometric function. In order to find the the bound-states solutions of the quantum system, this confluent hypergeometric function must be a finite degree polynomial $n$, and the quantity $\left(\beta+\frac{\gamma}{\chi}+\frac{1}{2}\right)$ should be a negative integer. This condition implies that $\left(\beta+\frac{\gamma}{\chi}+\frac{1}{2}\right)=-n$, where $n=0,1,2,...$. After simplifying this condition, one will find the following expression of the energy spectrum
\begin{equation}
E_{n,l}=\frac{1}{\alpha^2\,\Delta^2+e^2\,\kappa^2}\,\left[M\,e\,\kappa\,\eta \pm\,M \Delta\,\alpha^2\,\sqrt{\Delta^2+\frac{e^2\,\kappa^2-\eta^2}{\alpha^2}}\right],
\label{cc6}
\end{equation}
where 
\begin{equation}
\Delta=\left(n+\sqrt{\frac{2\,\xi\,(1-\alpha^2)+l'\,(l'+1)+\eta^2-e^2\,\kappa^2}{\alpha^2}+M\,\omega\,b+M^2\,\omega^2\,b^2}+\frac{1}{2}\right).
\label{cc7}
\end{equation}

\begin{figure}[ht]
\begin{subfigure}[b]{0.42\textwidth}
\includegraphics[width=3.2in,height=2.2in]{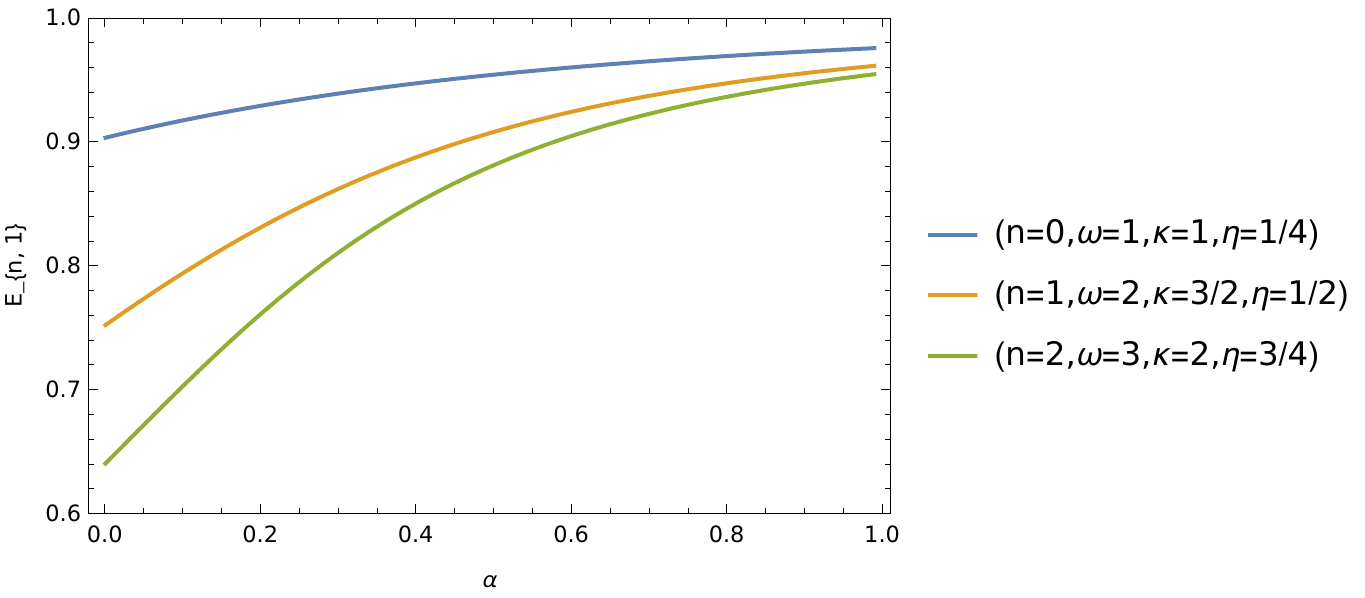}
\caption{$l=1, \Phi=1/4$}
\label{fig: 4 (a)}
\end{subfigure}
\hfill
\begin{subfigure}[b]{0.42\textwidth}
\includegraphics[width=3.2in,height=2.2in]{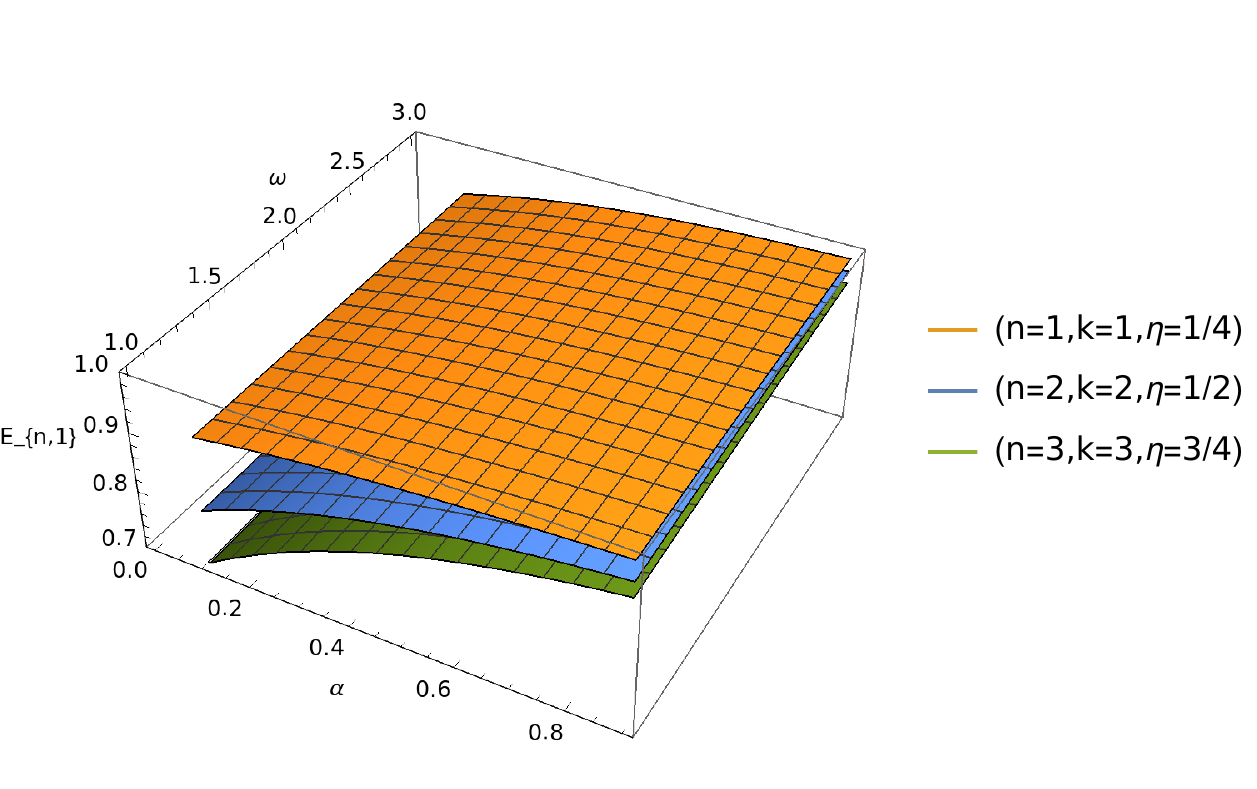}
\caption{$l=1, \Phi=1/4$}
\label{fig: 4 (b)}
\end{subfigure}
\hfill\\
\begin{subfigure}[b]{0.42\textwidth}
\includegraphics[width=3.2in,height=2.2in]{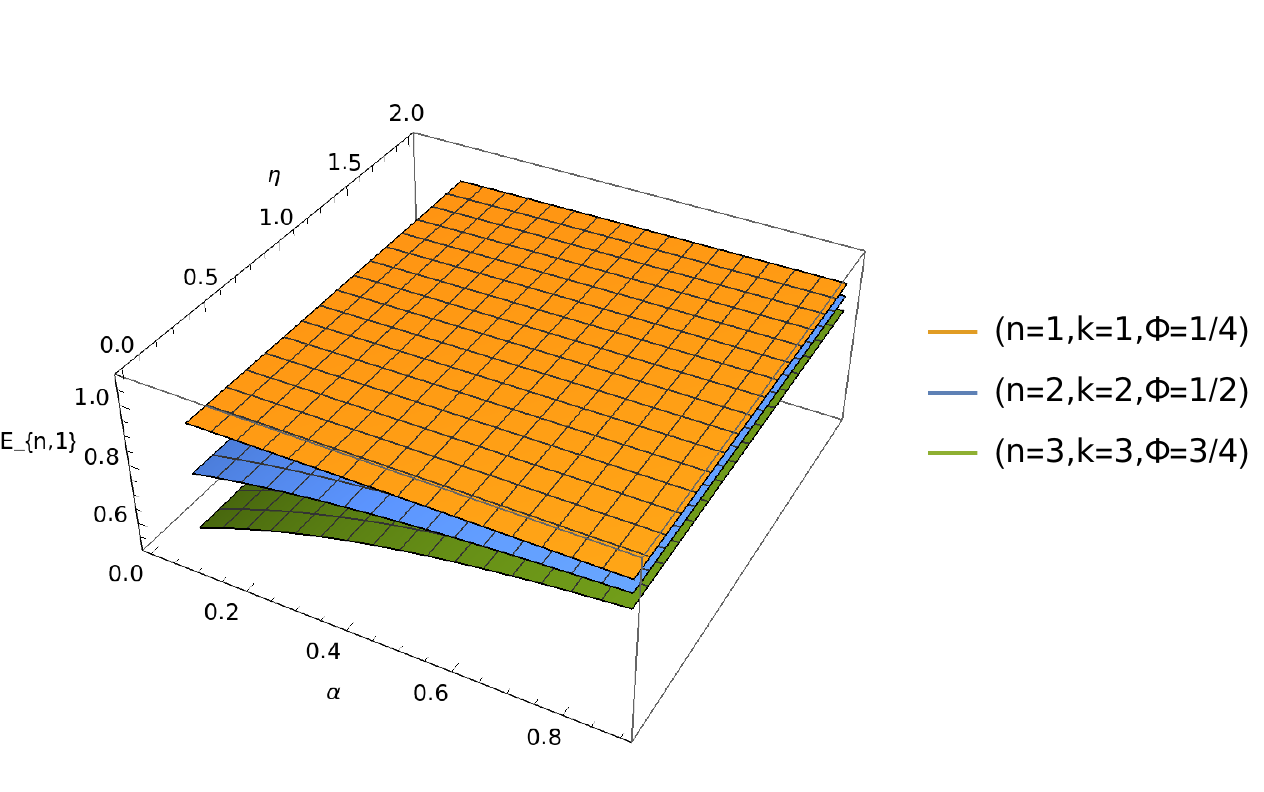}
\caption{$l=1, \omega=1$}
\label{fig: 4 (c)}
\end{subfigure}
\hfill
\begin{subfigure}[b]{0.42\textwidth}
\includegraphics[width=3.2in,height=2.2in]{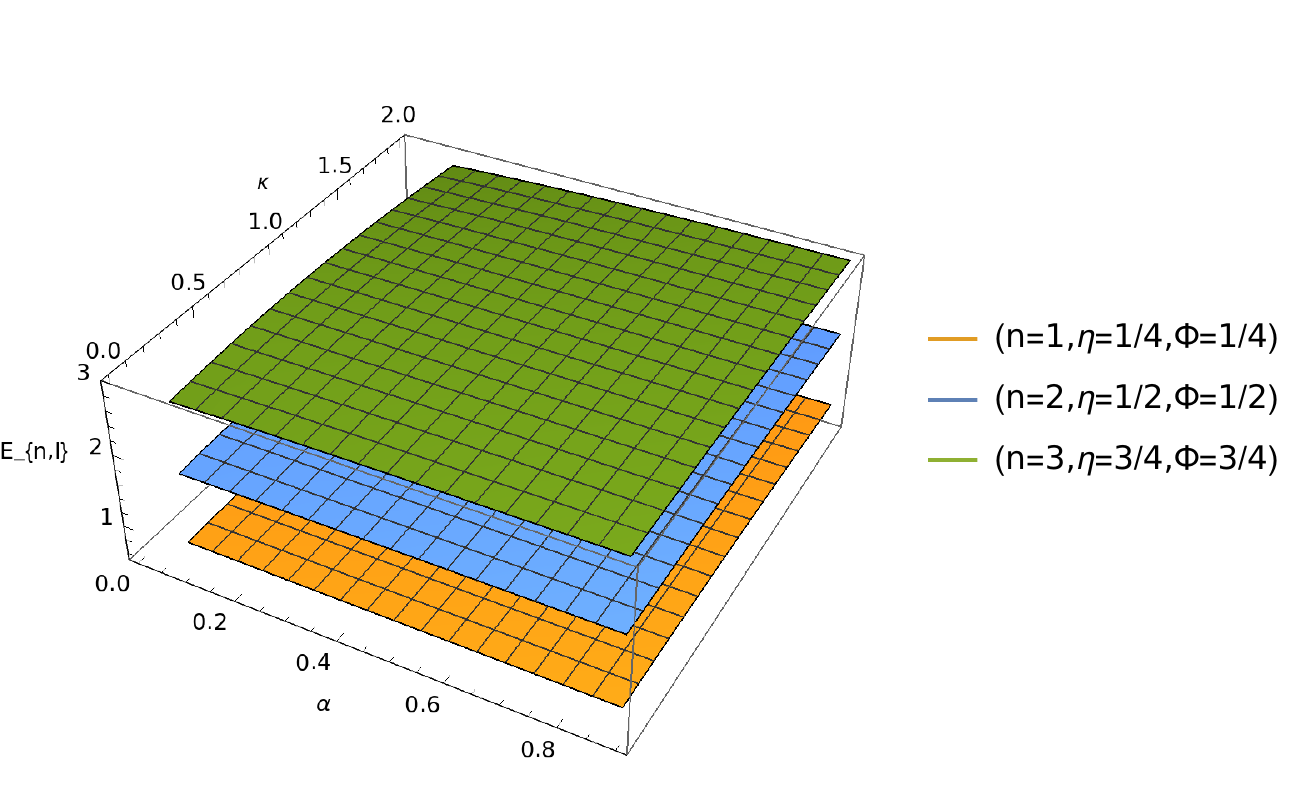}
\caption{$l=1, \omega=1$}
\label{fig: 4 (d)}
\end{subfigure}
\caption{Energy $E_{n,l}$ with parameters keeping fixed $M=1=b=e$, $\xi=1/2$ [units of different parameters are chosen in natural units $(c=1=\hbar)$].}
\label{fig: 4}
\end{figure}

The radial wave function is given by
\begin{eqnarray}
U_{n,l} (\rho)&=&\rho^{\beta}\,e^{-\frac{\rho}{2}}\,{}_1 F_{1} \Big(\beta+\frac{\gamma}{\chi}+\frac{1}{2}, 2\,\beta+1; \rho\Big),\nonumber\\
\psi_{n,l} (\rho)&=&\Big(2\,\chi\Big)^{1/2}\,\rho^{(\beta-\frac{1}{2})}\,e^{-\frac{\rho}{2}}\,{}_1 F_{1} \Big(\beta+\frac{\gamma}{\chi}+\frac{1}{2}, 2\,\beta+1; \rho\Big). 
\label{cc8}
\end{eqnarray}
We can see that the energy spectrum (\ref{cc6}) and the radial wave function (\ref{cc8}) of oscillator field are influenced by the topological defects parameter $\alpha^2$ which is associated with the curvature of the space-time, the function $f(r)=\frac{b}{r}$ as well as the Coulomb-types scalar and vector potentials present in the system. Furthermore, the energy eigenvalues $E_{n,l}$ depends on the Aharonov-Bohm magnetic flux $\Phi_B$ which gives us the gravitational analog of the Aharonov-Bohm effect \cite{YA,MP}. We have plotted graphs showing the effects of different parameters on the energy spectrum (fig. 4), and the wave function (fig. 5) of these oscillators field.

\begin{figure}[ht]
\begin{subfigure}[b]{0.45\textwidth}
\includegraphics[width=3.0in,height=2.0in]{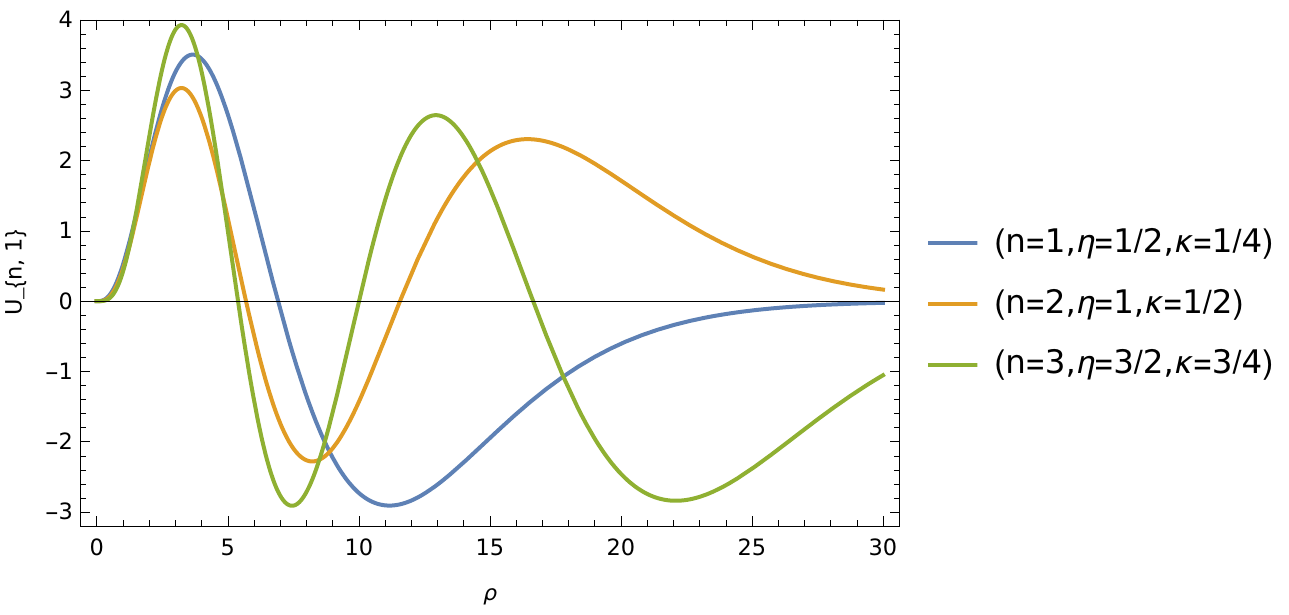}
\caption{$\Phi=1/2, \omega=1=l, \alpha=1/2$}
\label{fig: 5 (a)}
\end{subfigure}
\hfill
\begin{subfigure}[b]{0.45\textwidth}
\includegraphics[width=3.0in,height=2.0in]{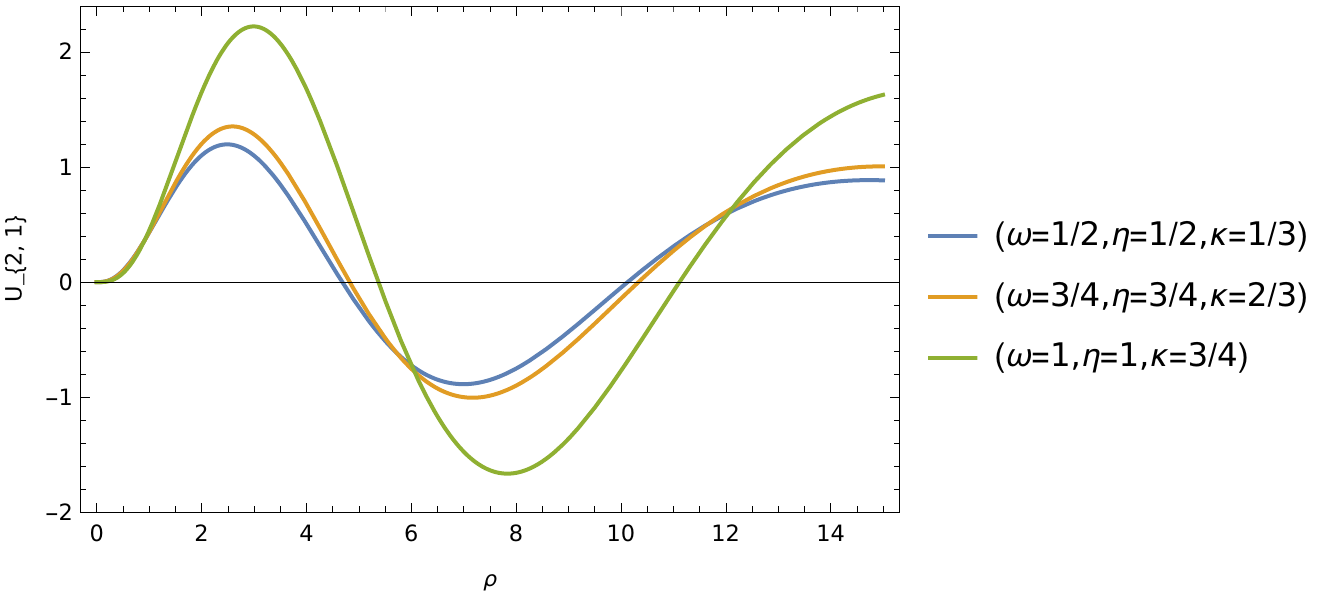}
\caption{$n=1=l, \Phi=1/2=\alpha$}
\label{fig: 5 (b)}
\end{subfigure}
\hfill\\
\begin{subfigure}[b]{0.45\textwidth}
\includegraphics[width=3.0in,height=2.0in]{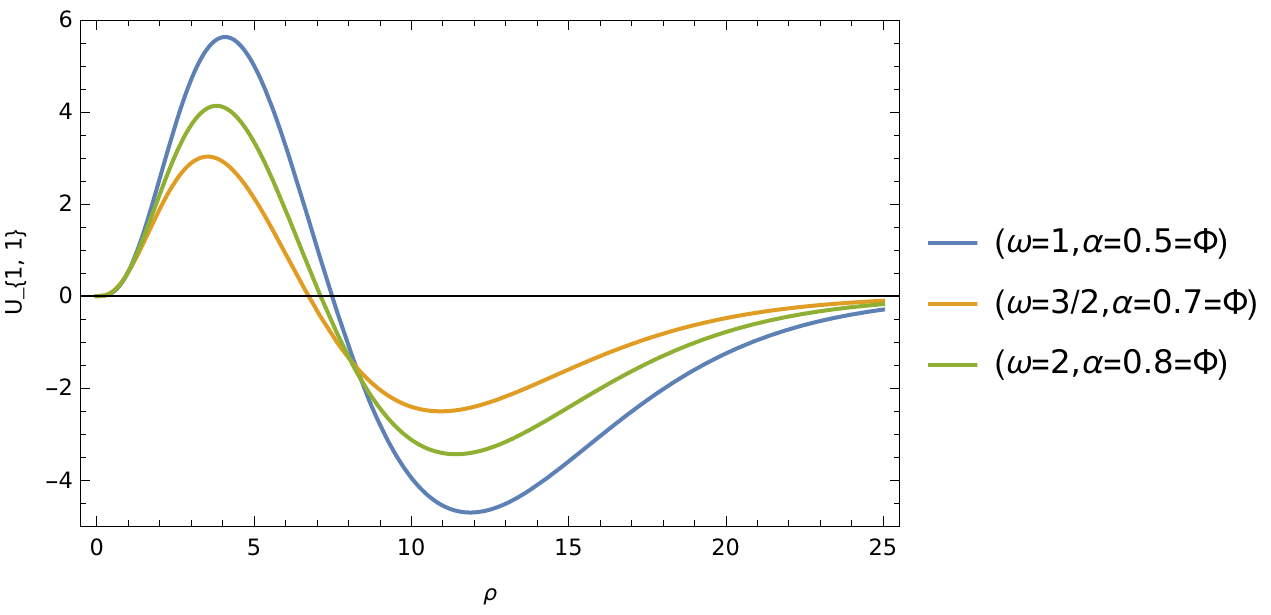}
\caption{$n=1=l, \eta=1/2, \kappa=1/4$}
\label{fig: 5 (c)}
\end{subfigure}
\hfill
\begin{subfigure}[b]{0.45\textwidth}
\includegraphics[width=3.4in,height=2.0in]{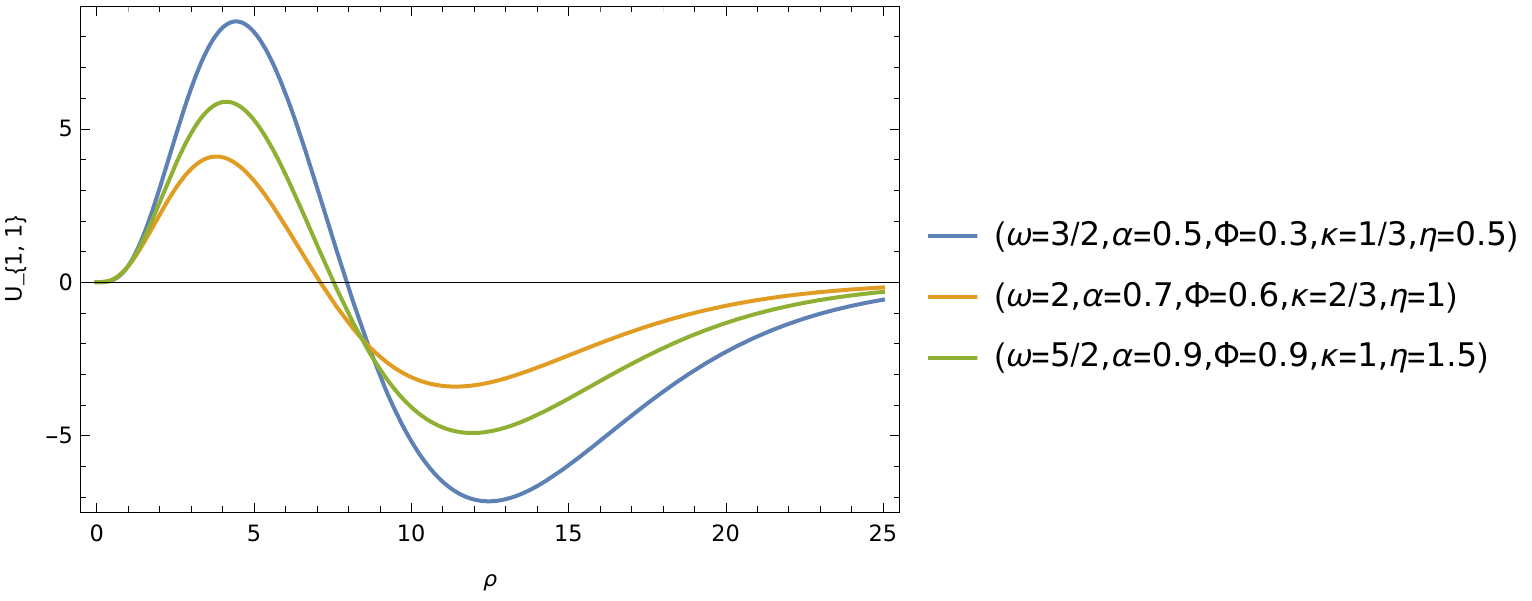}
\caption{$n=1=l$}
\label{fig: 5 (d)}
\end{subfigure}
\caption{Wave function $U_{n,l}$ keeping fixed $M=1=b=e$, $\xi=1/2$ [unit of different parameters are chosen in natural units $(c=1=\hbar)$].}
\label{fig: 5}
\end{figure}

\subsection{\bf Cornell-Type Function $f(r)=\left(a\,r+\frac{b}{r}\right)$ With Coulomb-Type Vector $A_0=\pm\,\frac{|\kappa|}{r}$ and Scalar $S=\frac{\eta}{r}$ Potentials.}

In this section, we study the generalized Klein-Gordon oscillator subject to a vector and scalar potentials of Coulomb-types (\ref{potential}) with a Cornell-type function, $f (r)=\left(a\,r+\frac{b}{r}\right)$ in the presence of an Aharonov-Bohm flux in a point-like global monopole space-time. We discuss the influences on the energy profiles of these oscillators. 

Thereby, substituting the Cornell-type function $f(r)=\left(a\,r+\frac{b}{r}\right)$ and Coulomb-type scalar and vector potential into the Eq. (\ref{9}), we have obtained the following radial wave equation:
\begin{equation}
\psi'' (r)+\frac{2}{r}\,\psi'(r)+\left(-\delta^2-\frac{\beta^2}{r^2}-\frac{2\,\gamma}{r}\right)\,\psi(r)=0,
\label{bb1}
\end{equation}
where we have defined
\begin{equation}
\delta^2=3\,M\,\omega\,a+2\,a\,b\,M^2\,\omega^2+\frac{1}{\alpha^2}\,(M^2-E^2).
\label{bb2}
\end{equation}

Now, we perform a transformation via $\psi (r) =\frac{U (r)}{\sqrt{r}}$ into the Eq. (\ref{bb1}), we have
\begin{equation}
U'' (r)+\frac{1}{r}\,U' (r)+\left(-\delta^2-\frac{\beta^2}{r^2}-\frac{2\,\gamma}{r} \right)\,U(r)=0.
\label{bb3}
\end{equation}
Considering a transformation of the radial coordinate via $\rho=2\,\delta\,r$ into the Eq. (\ref{bb3}), we have
\begin{equation}
U'' (\rho)+\frac{1}{\rho}\,U' (\rho)+\left(-\frac{\beta^2}{\rho^2}-\frac{\gamma}{\delta\,\rho}-\frac{1}{4} \right)\,U (\rho)=0.
\label{bb4}
\end{equation}
As stated earlier, the radial wave-function $U (\rho) \to 0$ for $\rho \to 0$ and $\rho \to \infty$. Suppose, a possible solution to the Eq. (\ref{bb4}) given by
\begin{equation}
U (\rho)=\rho^{\beta}\,e^{-\frac{\rho}{2}}\,F (\rho), 
\label{bb5}
\end{equation}
where $F (\rho)$ is an unknown function. Substituting this solution into the Eq. (\ref{bb4}), we have
\begin{equation}
\rho\,F''(\rho)+(2\,\beta+1-\rho)\,F'(\rho)+\left(-\beta-\frac{\gamma}{\delta}-\frac{1}{2}\right)\,F(\rho)=0.
\label{bb6}
\end{equation}
Equation (\ref{bb6}) is the well-known confluent hypergeometric equation which is a second order linear homogeneous differential equation. The solution of this Eq. (\ref{bb6}) that is regular for $\rho \to 0$ is given in terms of the confluent hypergeometric function as
\begin{equation}
F (\rho)= {}_1 F_{1} \Big(\beta+\frac{\gamma}{\delta}+\frac{1}{2}, 2\,\beta+1 ; \rho\Big).
\label{bb7}
\end{equation}
As state earlier, in order to have a bound-states solutions of this equation, the confluent hypergeometric function ${}_1 F_{1} (\beta+\frac{\gamma}{\delta}+\frac{1}{2}, 2\,\beta+1 ; \rho \to \infty)$ must be a finite degree polynomial of degree $n$, and the quantity $\left(\beta+\frac{\gamma}{\delta}+\frac{1}{2}\right)$ is a negative integer, that is, $\left(\beta+\frac{\gamma}{\delta}+\frac{1}{2}\right)=-n$, where $n=0,1,2,..$. After simplifying this condition, we have the following energy eigenvalues of the system
\begin{equation}
E_{n,l}=\frac{1}{\alpha^2\,\Delta^2+e^2\,\kappa^2}\,\left[M\,e\,\kappa\,\eta \pm\,M\,\Delta\,\alpha^2\, \sqrt{\Delta^2+\frac{e^2\,\kappa^2-\eta^2}{\alpha^2}+\Sigma^2}\right],
\label{bb8}
\end{equation}
where $\Delta$ is given earlier and
\begin{equation}
\Sigma=\sqrt{\omega\,a\,\left(\frac{3}{M}+2\,\omega\,b\right)\,(\alpha^2\,\Delta^2+e^2\,\kappa^2)}.
\label{bb9}
\end{equation}

The radial wave-function is given by
\begin{eqnarray}
U_{n,l} (\rho)&=&\rho^{\beta}\,e^{-\frac{\rho}{2}}\,{}_1 F_{1} \Big(\beta+\frac{\gamma}{\delta}+\frac{1}{2}, 2\,\beta+1; \rho\Big),\nonumber\\
\psi_{n,l} (\rho)&=&\Big(2\,\delta\Big)^{1/2}\,\rho^{(\beta-\frac{1}{2})}\,e^{-\frac{\rho}{2}}\,{}_1 F_{1} \Big(\beta+\frac{\gamma}{\delta}+\frac{1}{2}, 2\,\beta+1; \rho\Big). 
\label{bb10}
\end{eqnarray}

We can see that the energy spectrum (\ref{bb8})--(\ref{bb9}) of oscillator field are influenced by the topological defects parameter $\alpha^2$ of the space-time, the modified function $f(r)=\left(a\,r+\frac{b}{r}\right)$ as well as Coulomb-types scalar and vector potentials. Furthermore, the energy eigenvalues $E_{n,l}$ depends on the Aharonov-Bohm magnetic flux $\Phi_B$ which gives us the gravitational analog of the Aharonov-Bohm effect \cite{YA,MP}.

\section{Conclusions}

We have studied a spin-zero relativistic quantum oscillator (via the generalized Klein-Gordon oscillator) interacting with the gravitational field produced by the topological defects in a point-like global monopole (PGM) space-time. Later on, we insert scalar and vector potentials of Coulomb-types in the quantum system. We analyze the behaviour of the oscillator field under the topological defect that is associated with the scalar curvature of the space-time and see that the energy eigenvalues and the wave function are modified in comparison to the standard Landau levels. The present analysis can be used for simulation of a series of physical systems, for instance, the vibrational spectrum of diatomic molecules \cite{SMI}, binding of heavy quarks \cite{CQ,MC}, quark–antiquark interaction \cite{EE} etc.. The modified energy spectrum may be suitable to demonstrate the existence of these kinds of topological defects. However, from the observational point of view, it is clear that to have an observable modification in the energy spectrum of a physical system, we need a huge amount of particles unless the magnitude of this deviation from the original spectrum may not be strong enough to observe.

In this work, we have analyzed the effects of topological defects on a quantum oscillator field in the presence of external potentials. The radial wave equation of the generalized Klein-Gordon oscillator is derived. Then in {\tt sub-section 2.1}, we have chosen a Coulomb-type function $f(r)=\frac{b}{r}$ and considering zero vector and scalar potentials, $A_0=0=S$ into the radial equation, a second-order differential equation of the Bessel form is achieved. We solved this Bessel equation using the boundary condition of the hard-wall confining potential condition and obtained the relativistic energy eigenvalue by Eq. (\ref{16}). We have seen that the energy spectrum gets modified by the topology of the point-like global monopole space-time. In {\tt sub-section 2.2}, we have considered a Cornell-type function $f(r)=\Big(a\,r+\frac{b}{r}\Big)$ and considering zero vector and scalar potentials into the radial wave equation, the Whittaker differential equation \cite{KDM} form which is a second-order is achieved. After solving this Whittaker equation, the bound-states eigenvalues by the Eq. (\ref{aa7}) and the radial wave function by the Eq. (\ref{aa8}) of these oscillator fields are obtained. We have seen that the topological defects of the space-time geometry, as well as, the new function $f(r) \neq r$ modified the spectrum of energy and the wave function. In {\tt sub-section 2.3}, we have considered the Coulomb-type function $f(r)=\frac{b}{r}$ and Coulomb-type scalar and vector potentials (\ref{potential}) into the derived radial wave equation. We have achieved the confluent hypergeometric differential equation form and imposed the boundary condition for the bound-states solutions, one can find the bound-states energy eigenvalues by the Eq. (\ref{cc6}) and the radial wave function by the Eq. (\ref{cc8}) of these oscillator fields. One can see that the obtained energy eigenvalues, as well as the wave function, are modified by the topological defects and the Coulomb-type scalar and vector potentials considered in the quantum system. In {\tt sub-section 2.4}, we have considered Cornell-type function $f(r)=\Big(a\,r+\frac{b}{r} \Big)$ with Coulomb-types (\ref{potential}) vector and scalar potentials into the derived radial wave equation. Again we have achieved the confluent hypergeometric differential equation form, and following the previous procedure, we have obtained the bound-states energy eigenvalues by the Eq. (\ref{bb8}) and the radial wave function Eq. (\ref{bb10}) of these oscillator fields. Here also, we have seen that the energy spectrum and the wave unction are modified by the topological defects, the Coulomb-type scalar and vector potentials, and the considered function $f (r) \neq r$ on these oscillator fields.

An interesting feature of the presented results is that the energy profile of these oscillator fields in addition to the topological defect parameter $\alpha^2$ associated with the scalar curvature of the space-time and Coulomb-type potential depends on the Aharonov-Bohm magnetic flux $\Phi_B$. This is because the angular momentum quantum number $l$ is shifted, that is, $l \to l_{eff}=\left(l-\frac{e\,\Phi_B}{2\pi}\right)$, an effective angular momentum quantum number. Thus, the relativistic energy eigenvalue of the oscillator field is a periodic function of the geometric quantum phase $\Phi_B$, and we have that $E_{n,l} (\Phi_B \pm\,\Phi_0\,\tau)=E_{n,l\mp\tau} (\Phi_B)$, where $\tau=0,1,2,...$. This dependence of the energy eigenvalue on the geometric quantum phase gives us the gravitational analogue of the Aharonov-Bohm (AB) effect \cite{YA,MP}. Several authors have investigated this AB-effect in quantum mechanical systems Refs. \cite{EHR,AHEP2,EPJC,SR,IJGMMP,cc2,cc3,cc4}. It is well-known in condensed matter physics that the dependence of the energy eigenvalues on the geometric quantum phase gives rise to persistent currents that will discuss in future work.

Therefore, in this paper, we have shown some results for quantum systems where general relativistic effects are taken into account with the Aharonov-Bohm magnetic flux, which in addition with the previous results Refs. \cite{SZ,SZ3} present several interesting effects. This is a fundamental subject in physics, and the connections between these theories are not well understood.

\section*{Conflict of Interest}

There is no conflict of interests regarding publication of this paper.

\section*{Contribution}

F. Ahmed has done solely the whole work.

\section*{Funding Statement}

There is no funding agency associated with this manuscript.

\section*{Data Availability Statement}

All data generated or analysed during this study are included in this article [and its supplementary information files].

\section*{Acknowledgement}

Author would like to thank the anonymous kind referee(s) for their valuable comments and suggestion which have improved this paper a lot.

\end{document}